\documentclass[draftclsnofoot]{IEEEtran} \onecolumn
\onecolumn
\usepackage{setspace}
\usepackage[usenames]{color}
\usepackage{latexsym}
\usepackage{cite}
\usepackage{amssymb}
\usepackage{bm}
\usepackage{amsmath, amssymb}
\usepackage[dvips]{graphics}
\usepackage{graphicx}
\usepackage{psfrag}
 \allowdisplaybreaks

\begin{titlepage}
\title{{ Source Coding with a Side Information \\ ``Vending Machine''
}%
}
\author{
Haim Permuter\thanks{Haim Permuter is with the Department of
Electrical and Computer Engineering, Ben Gurion University of the
Negev, Beer-Sheva 84105, Israel. (Email: haimp@bgu.ac.il).} \and
Tsachy Weissman\thanks{Tsachy Weissman is with the Department of
Electrical Engineering, Technion, Haifa 32000,  Israel and on leave
from Department of Electrical Engineering, Stanford University,
Stanford, CA 94305, USA. (Email: tsachy@ee.technion.ac.il).} \ \ \
   }
\end{titlepage}

\begin{document}
\thispagestyle{empty} \setcounter{page}{1}
\setlength{\baselineskip}{1.3\baselineskip} \maketitle

\newtheorem{property}{Property}
\newtheorem{question}{Question}
\newtheorem{claim}{Claim}
\newtheorem{guess}{Conjecture}
\newtheorem{definition}{Definition}
\newtheorem{fact}{Fact}
\newtheorem{assumption}{Assumption}
\newtheorem{theorem}{Theorem}
\newtheorem{lemma}{Lemma}
\newtheorem{ctheorem}{Corrected Theorem}
\newtheorem{corollary}{Corollary}
\newtheorem{proposition}{Proposition}
\newtheorem{example}{Example}
\newtheorem{algorithm}{\underline{Algorithm}}[section]
\newcommand{\mat}[2]{\ensuremath{
\left( \begin{array}{c} #1 \\ #2 \end{array} \right)}}
\newcommand{\eq}[1]{(\ref{#1})}
\newcommand{\one}[1]{\ensuremath{\mathbf{1}_{#1}}}
\renewcommand{\thesubsection}{\thesection-\Alph{subsection}}
\newcommand{\am}{\mbox{argmin}}
\newcommand{\dmin}{d_{\mbox{min}}}
\newcommand{\be}{\begin{equation}}
\newcommand{\ee}{\end{equation}}
\newcommand{\eps}{\varepsilon}
\newcommand{\imipi}{\int_{-\infty}^{\infty}}
\newcommand{\mug}{\stackrel{\triangle}{=}}
\def \bfpi  {\bm{\pi}}
\def \bflambda  {\bm{\lambda}}
\def \bfdelta  {\bm{\delta}}

\begin{abstract}
We study source coding in the presence of side information, when the
system can take actions that affect the availability, quality, or
nature of the side information. We begin by extending the Wyner-Ziv
problem of source coding with decoder side information to the case
where  the decoder  is allowed to choose actions affecting the side
information. We then consider the setting where actions are taken by
the encoder, based on its observation of the source.
Actions may have costs that are commensurate with the quality of the
side information they yield, and  an overall per-symbol cost
constraint may be imposed. We characterize the achievable tradeoffs
between rate, distortion, and cost in some of these problem
settings. Among our findings is the fact that even in the absence of
a cost constraint, greedily choosing the action associated with the
`best' side information is, in general, sub-optimal. A few examples
are worked out.
\end{abstract}

\section{Introduction} \label{sec: Introduction}
The role and potential benefit of Side Information (S.I.) in
lossless and lossy data compression is a central theme in
information theory. In ways that are well understood for various
source coding systems, S.I.\ can be a valuable resource, resulting
in significant performance boosts relative to the case where it is
absent. In the problems studied thus far, the lack or availability
of the S.I., and its quality, are a given. But what if the system
can take actions that affect the availability, quality, or nature of
the S.I.?

For example, consider a source coding system where the S.I.\ is a
sequence of noisy measurements of the source sequence to be
compressed, each S.I.\ symbol acquired via a sensor. The quality of
each S.I.\ symbol may be commensurate with resources, such as power
or time expended by the sensor for obtaining it, which are limited.
Alternatively, or in addition, a sensor may have freedom to choose,
for each source symbol, how many independent noisy measurements to
observe, with a constraint on the overall number of measurements. It
is then natural to wonder how these resources, which may or may not
be limited, should best be used, and what would the corresponding
optimum performance be.
\begin{figure}[h!]{
\psfrag{X}[][][1]{$X^n$} \psfrag{Encoder}[][][1.1]{Encoder}
\psfrag{Decoder}[][][1.1]{Decoder} \psfrag{Vender}[][][1.1]{Vender}
\psfrag{T}[][][1]{$\;\;\;\;\;\;\;\;\;T(X^n)\in 2^{nR}$}
\psfrag{hat}[][][1]{$\;\hat X^n(Y^n,T)$}
\psfrag{T2}[][][1]{$A^n(T)\;\;\;\;\;$}
 \psfrag{Y}[][][1]{$Y^n$}
\psfrag{T5}[][][1]{$$} \psfrag{T6}[][][1]{$$}


\centerline{\includegraphics[width=10cm]{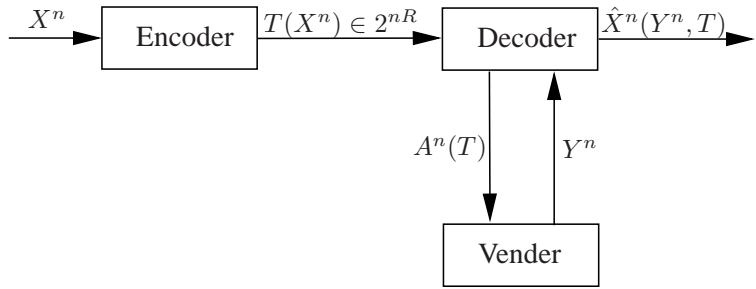}}

\caption{Rate distortion with a side information vender at the
decoder. The source $X^n$ is i.i.d. $\sim P_{X}$, and $Y^n$ is the
output of the side information channel $P_{Y|X,A}$ in response to
the pair of sequences $X^n$ and $A^n$, where $A^n$ is the action
sequence chosen by the decoder.} \label{f_decoder_vender}
}\end{figure}

We abstract this problem by assuming a memoryless source $P_X$,  a
conditional distribution of the side information given the source
\emph{and an action} $P_{Y|X,A}$,  a function  assigning costs to
the possible actions,  and a distortion measure. The first scenario
we focus on is that depicted in Figure \ref{f_decoder_vender}, where
the actions are taken at the decoder: Based on its observation of
the source
  sequence $X^n$, which is i.i.d.$\sim P_X$, the encoder gives an index  to  the
  decoder. Having received the
  index, the decoder chooses the action
  sequence $A^n$. Nature then generates the side information sequence
  $Y^n$ as the output of the memoryless channel $P_{Y|X,A}$ whose
  input is the pair $(X^n, A^n)$. The reconstruction sequence
  $\hat{X}^n$ is then based on the index and on the side
  information sequence.

The setting of Figure \ref{f_decoder_vender} can be considered the
source coding dual of coding for channels with action-dependent
states, where  the transmitter chooses an action sequence that
affects the formation of the channel states, and  then creates the
channel input sequence based on the state sequence, as considered in
\cite{verduweissmanisit09}.
  We characterize the
achievable tradeoff between rate, distortion, and cost in Section
\ref{sec: Side Information Vending Machine at the Decoder}. We
demonstrate, by a few examples, that  greedily choosing the action
associated with the `best' side information may be sub-optimal even
in the absence of a cost constraint. Further, in the presence of a
cost constraint, time-sharing between schemes that are optimal for
different cost values is, in general, sub-optimal. We also
characterize the fundamental limits for the case where the
reconstruction is confined to causal dependence on the side
information sequence, and the case where the encoder observes a
noisy observation of the source rather than the source itself.

\begin{figure}[h!]{
\psfrag{X}[][][1]{$X^n$} \psfrag{Encoder}[][][1.1]{Encoder}
\psfrag{Decoder}[][][1.1]{Decoder} \psfrag{Vender}[][][1.1]{Vender}
\psfrag{T}[][][1]{$\;\;\;\;\;\;\;\;\;T\in 2^{nR}$}
\psfrag{hat}[][][1]{$\;\hat X^n(Y^n,T)$}
\psfrag{T2}[][][1]{$A^n(X^n)\;\;\;\;\;\;\;\;$}
 \psfrag{Y}[][][1]{$Y^n$}
\psfrag{T5}[][][1]{$$} \psfrag{T6}[][][1]{$$}
\centerline{\includegraphics[width=10cm]{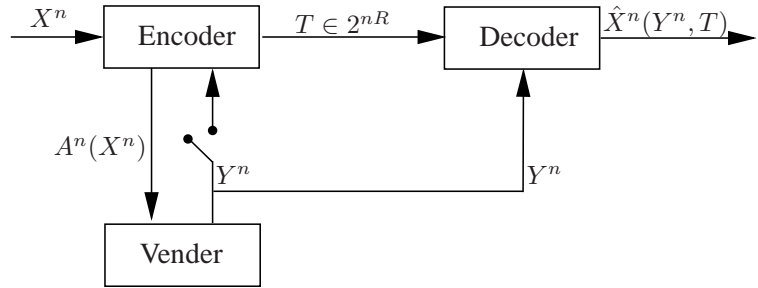}}
\caption{Rate distortion with side information vender at the
encoder, where the side information is known at the decoder and may
or may not be known to the encoder. The source $X^n$ is i.i.d.$\sim
P_X$ and side information is generated as the output of the
memoryless channel $P_{Y|X,A}$ in response to the input $(X^n,
A^n)$, where the action sequence $A^n$ is chosen by the encoder.}
\label{f_action_encoder_switch} }\end{figure}

The second scenario we consider is that depicted in Figure
\ref{f_action_encoder_switch}, where actions are taken at the
encoder: Based on its observation of the source
  sequence $X^n$,  the encoder chooses a sequence of actions $A^n$. Nature then generates the side information sequence
  $Y^n$ as the output of the memoryless channel $P_{Y|X,A}$ whose
  input is the pair $(X^n, A^n)$. The encoder now chooses the index  to be given to the
  decoder on the basis of the source and possibly the side information
  sequence (according to whether or not the switch is closed).   The reconstruction sequence
  $\hat{X}^n$ is then based on the index and on the side
  information sequence.
%
%
   Though we leave the
general case open, in Section \ref{sec: Side Information Vending
Machine at the Encoder} we characterize the  achievable tradeoff
between rate, distortion, and cost for three important special
cases: the (near) lossless case, the Gaussian case (where $Y=A+X+N$,
with $X$ and $N$ being independent Gaussian random variables), and
the case of the Markov relation $Y-A-X$ (i.e., when $P_{Y|X,A}$ is
of the form $P_{Y|A}$). We end that section with Subsection
\ref{subsec: Upper and  Lower  Bounds  for the General Case}, giving
lower and upper bounds on the achievable rates for the general case.
We summarize the paper and related open directions in Section
\ref{sec: Summary and Open Questions}.


The family of problems we consider in this work includes scenarios
arising naturally in the coding or compression of sources for which
the S.I.\ arises from noisy measurements of the source components.
The acquisition, handling, processing  and storage of these
measurements may require system resources that come at a cost. This
premise, that the acquisition of source measurements may be costly
and is to be done sparingly, is in fact central in the emerging
Compressed Sensing paradigm \cite{CandesRombergTao2006,
 CandesTao2006, Donohocompressedsensing}, arising naturally in the study
of an increasing array of sensing problems. In many such problems,
the system has the freedom to choose how many sensors to deploy in
each region of the phenomenon it is trying to gauge, subject to an
overall budget of sensors. Assuming each sensor provides an
independent measurement of the source region in which it was
deployed, this setting corresponds to our model, with $A_i \in \{ 0,
1, 2, \ldots \}$ representing the number of sensors, $P_{Y|X,A} =
\prod_{j=1}^A P_{Z_j |X}$ representing $A$ independent measurements
from the `sensor channel' $P_{Z|X}$, and $\Lambda(A_i) = A_i$
assuming all sensors are equally costly. The cost constraint $C$
then corresponds to the budget of sensors to deploy, in number of
sensors per source region. We are not  aware of previous work on
source coding for systems allowed to take S.I.-affecting actions
from a Shannon theoretic perspective. We refer to
\cite{MartinianWornellZamir} and some references therein for other
recent Shannon theoretic studies of new problems involving source
coding in the presence of S.I.


\section{Side Information Vending Machine at the Decoder}
\label{sec: Side Information Vending Machine at the Decoder}

Throughout the paper we let upper case, lower case, and calligraphic
letters denote, respectively, random variables, specific or
deterministic values they may assume, and their alphabets. For two
jointly distributed random objects $X$ and $Y$, let $P_X$,
$P_{X,Y}$, and $P_{X|Y}$ respectively denote the distribution of
$X$, the joint distribution of $X,Y$, and the conditional
distribution of $X$ given $Y$. In particular, when $X$ and $Y$ are
discrete, $P_{X|Y}$ represents the stochastic matrix whose elements
are $P_{X|Y} (x|y) = P(X=x|Y=y)$. The term $X_m^n$ denotes the
$n-m+1$-tuple $(X_m, \ldots, X_n)$ when $m \leq n$ and the empty set
otherwise. The term $X^n$ is shorthand for $X_1^n$, and $X^{n
\setminus i}$ stands for the $n-1$-tuple consisting of all the
components of $X^n$ but $X_i$.

\subsection{The Setup} \label{sec: the setup} A source with action
dependent decoder side information is characterized by the source
distribution $P_{X}$ and by the conditional distribution of the side
information given the source \underline{and an action} $P_{Y|X,A}$.
The difference between this and previously studied scenarios is that
here, after receiving the index from the encoder, the decoder may
choose actions that will affect the nature of the side information
it will get to observe. Specifically, a scheme in this setting for
blocklength $n$ and rate $R$ is characterized by an encoding
function $T : \mathcal{X}^n \rightarrow \{1,2, \ldots, 2^{nR} \}$,
an action strategy $f: \{1,2, \ldots, 2^{nR} \} \rightarrow
\mathcal{A}^n$, and a decoding function $g : \{1,2, \ldots, 2^{nR}
\} \times \mathcal{Y}^n \rightarrow \hat{\mathcal{X}}^n$ that
operate as follows:
\begin{itemize}
  \item The source $n$-tuple $X^n$ is i.i.d. $\sim P_X$
  \item Encoding: based on $X^n$ give index $T = T(X^n)$ to the decoder

  \item Decoding:
  \begin{itemize}
    \item given the index, choose an action sequence $A^n = f(T)$

    \item   the side information
    $Y^n$ will be the output of the memoryless channel $P_{Y|X,A}$
    whose input is $(X^n, A^n)$

    \item let $\hat{X}^n = g (T, Y^n)$
  \end{itemize}
\end{itemize}

A triple $(R,D,C)$ is said to be \emph{achievable} if for all $\eps
>0$ and sufficiently large $n$ there exists a scheme as above for blocklength $n$ and rate
$R + \eps$ satisfying both
\begin{equation}\label{eq: distortion constraint}
    E \left[ \sum_{i=1}^n \rho (X_i, \hat{X}_i) \right] \leq n (D+\eps)
\end{equation}
and
\begin{equation}\label{eq: cost constraint}
    E \left[ \sum_{i=1}^n \Lambda (A_i) \right] \leq n (C+\eps) ,
\end{equation}
where $\rho$ and $\Lambda$ are, respectively, given distortion and
cost functions. The rate distortion (and cost) function $R(D,C)$ is
defined as
\begin{equation}\label{eq: rate distortion function defined operationally}
R(D,C) = \inf \{ R' : \mbox{ the triple } (R',D,C) \mbox{ is
achievable} \}.
\end{equation}

\subsection{The Rate Distortion Cost Tradeoff} Define
\begin{equation}\label{eq: rd function for action dependent si}
    R^{(I)}(D, C) = \min \left[ I(X ; A) + I(X ; U | Y, A) \right],
\end{equation}
where the joint distribution of $X,A,Y,U$ in \eq{eq: rd function for
action dependent si} is of the form
\begin{equation}\label{eq: form of joint dist for act-dep rd func}
  P_{X,A,U,Y}(x,a,u,y) =  P_X (x) P_{A,U|X} (a,u|x)
  P_{Y|X,A}(y|x,a),
\end{equation}
and the minimization is over all $P_{A,U|X}$ under which
\begin{equation}\label{eq: the expected distortion constraint}
    E \left[ \rho \left( X, \hat{X}^{opt} (U,Y) \right) \right] \leq
    D,  \ \ \ \ E \left[  \Lambda (A) \right]
\leq C,
\end{equation}
where $\hat{X}^{opt} (U,Y)$ denotes the best estimate of $X$ based
on $U,Y$, $U$ is an auxiliary random variable.
We show below that the cardinality of $U$ may be restricted to
$|\mathcal{U}| \leq |\mathcal X||\mathcal A|+1$. Our main result
pertaining to $R^{(I)}(D, C)$ is the following:
\begin{theorem} \label{th: main theorem for si vending at decoder}
The rate distortion cost function, as defined in \eq{eq: rate
distortion function defined operationally}, is given by $R^{(I)}(D,
C)$ in \eq{eq: rd function for action dependent si}, i.e.,
\begin{equation}\label{eq: rdc func equals operational}
R(D,C) =   R^{(I)}(D, C).
\end{equation}
\end{theorem}
Remark: Write $R_{WZ} (P_X, P_{Y|X}, D)$ for the explicit dependence
of the Wyner-Ziv rate distortion function \cite{WynerZiv1976} on the
distribution of the source and the conditional distribution of the
source given the side information. It is clear that
\begin{equation}\label{eq: can always time-share between WZ coding}
R(D,C) \leq  \min \left\{ \sum_a P_{A}(a) R_{WZ} (P_X, P_{Y|X, A=a},
D_a) :  \sum_a P_{A}(a) D_a \leq D  , \sum_a P_{A}(a) \Lambda (a)
\leq C \right\},
\end{equation}
since the right hand side can be achieved by letting the decoder
take actions according to a pre-specified sequence with the symbol
$a$ fraction $P_A (a)$ of the time, and performing Wyner-Ziv coding
at distortion level $D_a$ separately on each subsequence associated
with each action symbol. It is natural to wonder whether the
inequality in \eq{eq: can always time-share between WZ coding} can
be strict. We will see through some examples below that, in general,
it may very well be strict. Indeed, even in the absence of a cost
constraint, we give examples showing that greedily selecting the
action associated with the side information which is best in the
Wyner-Ziv sense, that is the action $a$ minimizing $R_{WZ} (P_X,
P_{Y|X, A=a}, D)$, may be suboptimal.

The following lemma will be useful in proving Theorem \ref{th: main
theorem for si vending at decoder}.
\begin{lemma}{\it  Properties of the expressions defining $R^{(I)}(D,
C)$:\label{lemma:properties_R}}

\begin{enumerate}
\item \label{lemma_properties_R_convex}
For any fixed $P_X$ and $P_{Y|A,X}$, the set of distributions of the
form given in (\ref{eq: form of joint dist for act-dep rd func}) is
a convex set in $P_{A,U|X}$.

\item
For any fixed $P_X$ and $P_{Y|A,X}$, the expression $I(X ; A) + I(X
; U | Y, A)$ is convex in $P_{A,U|X}$ (assuming the joint
distribution given in (\ref{eq: form of joint dist for act-dep rd
func})).
\item \label{lemma_properties_R_size}
To exhaust  $R^{(I)}(D, C)$, it is enough to restrict the alphabet
of $U$ to satisfy
\begin{eqnarray}
|{\cal U}|\leq |\mathcal X||\mathcal A|+2.
\end{eqnarray}

\item It suffices to restrict the minimization in \eq{eq: rd
function for action dependent si} to joint distributions where $A$
is a deterministic function of $U$, i.e., of the form
\begin{equation}\label{eq: more restricted form of the joint dist}
    P_X (x) P_{U|X} (u|x) 1_{\{ a = f(u) \}}
  P_{Y|X,A}(y|x,a).
\end{equation}
\end{enumerate}
\end{lemma}
\emph{Proof:}
\begin{enumerate}
\item Since the set of conditional distributions $P_{A,U|X}$ is a convex
set, and since $P_X$ and $P_{Y|X,A}$ are fixed, the set of
distributions $P_{X,A,U,Y}$ of the form given in (\ref{eq: form of
joint dist for act-dep rd func}) is a convex set.\hfill \QED

\item \label{it: convexity}Using the definition of mutual information we have the identity,
\begin{equation}\label{eq: simple region identity}
I(X ; A) + I(X ; U | Y, A)=I(X;U,Y,A)+H(Y|A,X)-H(Y|A).
\end{equation}
We show now that the right-hand part of (\ref{eq: simple region
identity}) is convex in $P_{U,A|X}$ for a fixed $P_X$ and
$P_{Y|A,X}$. The expression $I(X;U,Y,A)$ is convex in $P_{U,Y,A|X}$,
hence it is also convex in $P_{U,A|X}$. For fixed $P_X$ and
$P_{Y|A,X}$, the expression $H(Y|A,X)$ is linear in $P_{A|X}$.
Finally, we show that $-H(Y|A)$ is convex using the the log sum
inequality that states that for non negative number, $a_1,a_2$ and
$b_1,b_2$
\begin{equation}
a_1 \log \frac{a_1}{b_1}+a_2 \log \frac{a_2}{b_2}\geq \left(a_1+a_2
\right) \log \frac{a_1+a_1}{b_1+b_2}.
\end{equation}
Now let $P^{3}_{A|X}=\alpha P^{1}_{A|X} + \overline{\alpha}
P^{2}_{A|X}$, where $0\leq \alpha\leq 1$ and
$\overline{\alpha}=1-\alpha$. Let us denote $P^i_{Y,A,X}$ and
$H^{i}(A|X)$, the joint distribution and the conditional entropy
induced by $P^{i}_{A|X}$ and the fixed pmfs $P_X$ and $P_{Y|A,X}$
for $i=1,2,3$. Consider,
\begin{eqnarray}\label{eq: ineq1}
P^3_{Y,A}(y,a)\log\frac{P^3_{Y,A}(y,a)}{P^3_{A}(a)}&\stackrel{(a)}{=}&
\left (\alpha P^1_{Y,A}(y,a)+\overline \alpha P^2_{Y,A}(y,a)\right
)\log\frac{\alpha P^1_{Y,A}(y,a)+\overline \alpha
P^2_{Y,A}(y,a)}{\alpha P^1_{A}(a)+\overline \alpha
P^2_{A}(a)}\nonumber \\
&\stackrel{(b)}{\leq}& \alpha P^1_{Y,A}(y,a)\log\frac{
P^1_{Y,A}(y,a)}{ P^1_{A}(a)}+\overline \alpha
P^2_{Y,A}(y,a)\log\frac{ P^2_{Y,A}(y,a)}{ P^2_{A}(a)},
\end{eqnarray}
where (a) follows from the definition of $P^i_{Y,A,X}$ and (b)
follows from the log sum inequality. Since (\ref{eq: ineq1}) holds
for any $a\in \mathcal A$ and any $y\in \mathcal Y$, we obtain that
$-H(Y|A)$ is convex, i.e.,
\begin{equation}
-H^3(Y|A)\leq -\alpha H^1(Y|A)-\overline \alpha H^2(Y|A)
\end{equation}
\hfill \QED

\item \label{it: cardinality} We invoke the
support lemma~\cite{Csiszar81}. The external random variable $U$
must have $|{\cal X}||{\cal A}|-1$ letters to preserve $P_{X,A}$,
plus two more to preserve the distortion constraint, the cost
constraint  and $I(A;X)+I(X ; U | Y, A)$. This results in alphabet
of size $|{\cal X}| |{\cal A}|+2$.\hfill \QED

\item Note that it suffices to restrict the minimization in \eq{eq: rd
function for action dependent si} to joint distributions where $A$
is a deterministic function of $U$, i.e., of the form
\begin{equation}\label{eq: more restricted form of the joint dist again bef}
    P_X (x) P_{U|X} (u|x) 1_{\{ a = f(u) \}}
  P_{Y|X,A}(y|x,a),
\end{equation}
in lieu of \eq{eq: form of joint dist for act-dep rd func}. To see
the equivalence note that a distribution of the form in \eq{eq: form
of joint dist for act-dep rd func} assumes the form in \eq{eq: more
restricted form of the joint dist} by taking $(U,A)$ as the
auxiliary variable.\hfill \QED
\end{enumerate}
\emph{Proof  of Theorem \ref{th: main theorem for si vending at
decoder}:}

{\bf Achievability:} We briefly and informally outline the
achievability part, which is based on standard arguments: A
code-book of size $2^{n (I(X;A) + \eps)}$ is generated with
codewords that are i.i.d.$\sim P_A$. For each such codeword,
generate $2^{n (I(X;U|A) + \eps)}$ codewords according to $P_{U|A}$.
Distribute these codewords uniformly at random into $2^{n
(I(X;U|Y,A) + 2 \eps)}$ bins.   Given the source realization, $n
(I(X;A) + \eps)$ bits are used by the encoder to communicate the
identity of a codeword from the first codebook jointly typical with
it (with high probability there is at least one such codeword). The
decoder now performs the actions according to the action sequence
conveyed to it. The encoder now uses an additional $n (I(X;U|Y,A) +
2 \eps)$ number of bits to describe the bin index of the codeword
from the second code-book which is jointly typical with the source
and the first codeword. With high probability there is at least one
such codeword (since more than $2^{n I(X;U|A)}$ such were
generated), and it is the only codeword in its bin which is jointly
typical with the first codeword (which the decoder already knows)
and the side information sequence that it has generated and is
observed at the  decoder, since the size of each bin is no larger
than $\approx 2^{n (I(X;U|A) - I(X;U|Y,A) - \eps)} = 2^{n (I(Y;U|A)
- \eps)}$. For the reconstruction, the decoder now employs the
mapping $\hat{X}^{opt}$ in a symbol-by-symbol fashion on the
components of the pair consisting of the second codeword and the
side information sequence.

{\bf Converse:} For the converse part, fix a scheme of rate $\leq R$
for a block of length $n$ and consider:
\begin{eqnarray}
    n R & \geq & H(T) \nonumber \\ & = & H(T, A^n)  \nonumber \\ & = & H(A^n) + H(T | A^n ) \nonumber  \\ & \geq &
    H(A^n) - H(A^n | X^n) + H(T|A^n, Y^n) - H(T|Y^n, A^n, X^n)  \nonumber \\ &
    = & I(X^n ; A^n) + I(X^n; T|A^n, Y^n)  \nonumber \\ & = &  I(X^n ; A^n) + H(X^n|A^n,
    Y^n) - H(X^n|A^n,Y^n,T). \label{eq: converse chain of ineq}
\end{eqnarray}
Now
\begin{eqnarray}
   I(X^n ; A^n) + H(X^n|A^n, Y^n) & \geq & H(X^n) - H(X^n|A^n) + H(X^n, Y^n|A^n) - H(Y^n|A^n) \nonumber  \\
    & = & H(X^n) - H(X^n|A^n) + H(X^n|A^n) + H(Y^n|A^n, X^n) -
    H(Y^n|A^n)   \nonumber \\   & = & H(X^n) + H(Y^n|A^n, X^n) - H(Y^n|A^n)
    \nonumber \\   & = & \sum_{i=1}^n H(X_i) + H(Y_i|A_i, X_i) - H(Y_i| Y^{i-1},
    A^n) \label{eq: follows by breaking up with no fb}
      \\   & \geq & \sum_{i=1}^n H(X_i) + H(Y_i|A_i, X_i) - H(Y_i|  A_i )
     \nonumber \\   & = & \sum_{i=1}^n H(X_i) - I(Y_i ; X_i|A_i)
    \nonumber  \\   & = & \sum_{i=1}^n I(X_i; A_i) + H (X_i| Y_i, A_i).
    \label{eq: converse chain of ineq part2}
\end{eqnarray}
Combining \eq{eq: converse chain of ineq} and \eq{eq: converse chain
of ineq part2} yields
\begin{eqnarray}
  n R & \geq & \sum_{i=1}^n I(X_i; A_i) + H (X_i| Y_i, A_i) - H(X_i|X^{i-1}, A^n, Y^n, T)\nonumber  \\
    &\stackrel{(a)}{=}& \sum_{i=1}^n I(X_i; A_i) + H (X_i| Y_i, A_i) - H(X_i|Y_i, A_i, U_i) \nonumber  \\
    &=&  \sum_{i=1}^n I(X_i; A_i) + I (X_i ; U_i | Y_i, A_i)  \label{eq: defining aux U},
\end{eqnarray}
where (a) follows by taking $U_i=(A^{n \setminus
i}, Y^{n \setminus i}, X^{i-1},T)$. Noting that $\hat{X}_i =
\hat{X}_i (T, Y^n)$ is a function of the pair $(U_i,Y_i)$, and the
Markov relation $U_i - (A_i, X_i)-Y_i$, the proof is now completed
in the standard way upon considering the joint distribution of
$(X',A',U', Y', \hat{X}') \mug (X_J, A_J, U_J, Y_J, \hat{X}_J)$,
where $J$ is randomly generated uniformly at random from the set
$\{1, \ldots, n\}$, independent of $(X^n, A^n, U^n, Y^n,
\hat{X}^n)$, and noting that:

  \begin{equation}\label{eq: distribution properties}
   P_{X'} = P_X,\ U' - (A',X')-Y',\ P_{Y'|X',A'} =
  P_{Y|X,A},
  \end{equation}
  \begin{equation}
   \hat{X}' = \hat{X}' (U', Y'),
  \end{equation}
\begin{equation}\label{eq: distortion and cost constraint single let}
    E \left[ \sum_{i=1}^n \rho (X_i, \hat{X}_i) \right] = n E \rho
    (X', \hat{X}')
, \ \ \ \
    E \left[ \sum_{i=1}^n \Lambda (A_i) \right] = n E \Lambda (A')
\end{equation}
   and
\begin{equation}\label{eq: convexity of mut inf for sum vs generic}
     \frac{1}{n} \sum_{i=1}^n I(X_i; A_i) + I (X_i ; U_i | Y_i, A_i) \geq I(X'; A') + I (X' ; U' | Y',
     A') ,
\end{equation}
where last inequality follows from item \ref{it: convexity} in Lemma
\ref{lemma:properties_R}, which states that $I(X; A) + I (X ; U | Y,
A)$ is convex over the set of distributions that satisfies (\ref{eq:
distribution properties}).\hfill \QED

It is natural to wonder whether the characterization above remains
valid when the choice of the actions is allowed to depend on the
side information symbols generated thus far, that is, for the $i$th
action to be of the form $A_i = A_i (T, Y^{i-1})$. The converse in
the proof above does not carry over to this case since the
inequality $H(Y^n|A^n, X^n) \geq \sum_{i=1}^n  H(Y_i|A_i, X_i)$,
used in \eq{eq: follows by breaking up with no fb}, may no longer
hold. Whether the best achievable rate could, in general, be better
(less) when allowing such  schemes remains open.


\subsection{Actions taken by the decoder before the index is seen}
Consider the setting as in Figure \ref{f_decoder_vender}, where the
actions $A^n$ are taken by the decoder before the index $T$ is seen.
In such a case $A^n$ is independent of $X^n$. For this case, the
rate distortion cost function is similar to $R^{(I)}(D,C)$  defined
in the previous section, but with an additional constraint that $A$
is independent of $X$. Define
\begin{equation}\label{eq: rd perp}
    R^{(I)}_{A \perp  X} (D, C) = \min  I(X ; U | Y, A) ,
\end{equation}
where the joint distribution of $X,A,Y,U$  is of the form
\begin{equation}\label{eq: form of joint dist for act-dep rd func againbef}
  P_{X,A,U,Y}(x,a,u,y) =  P_X (x) P_{A}(a) P_{U|X,A} (u|x,a)
  P_{Y|X,A}(y|x,a),
\end{equation}
and the minimization is over all $P_{A}$ and $P_{U|X,A}$ under which
\begin{equation}\label{eq: the expected distortion constraint}
    E \left[ \rho \left( X, \hat{X}^{opt} (U,Y) \right) \right] \leq
    D,  \ \ \ \ E \left[  \Lambda (A) \right]
\leq C,
\end{equation}
where $\hat{X}^{opt} (U,Y)$ denotes the best estimate of $X$ based
on $U,Y$, where $U$ is an auxiliary random variable with a
cardinality $|\mathcal{U}| \leq |\mathcal X||\mathcal A|+2$.

\begin{theorem} \label{th: actions before decoder}
The rate distortion cost function for the setting where actions
taken by the decoder before the index is seen, is given by
$R^{(I)}_{A \perp  X}(D, C)$.
\end{theorem}
\begin{proof}
The proof is similar to the proof of Theorem \ref{th: main theorem
for si vending at decoder}, but taking into account that $A^n$ is
independent of $X^n$, and therefore $A_i$ is independent of $X_i$.
\end{proof}

If the cost is unlimited, then the greedy policy is optimal, namely
the decoder blindly chooses the action $a$ minimizing
\begin{equation}
R_{WZ} (P_X, P_{Y|X, A=a}, D),
\end{equation}
and an optimal Wyner-Ziv code for the source $P_X$ and channel
$P_{Y|X, A=a}$ is employed. For the more general case, in the
presence of a cost constraint, as can be expected and  is
straightforward to check,
    $R^{(I)}_{A \perp  X} (D, C)$ in \eq{eq: rd perp} coincides with
    the minimum on the right hand side of \eq{eq: can always time-share between WZ
    coding}.

\subsection{Examples}
\subsubsection{The Lossless Case} As a very special case of Theorem
\ref{th: main theorem for si vending at decoder} we get that, in the
absence of a cost constraint on the actions, the minimum rate needed
for a near lossless reconstruction at  the decoder is  given by
\begin{equation}\label{eq: near lossless reconstrcution min rate}
    \min I(X ; A) + H(X  | Y, A),
\end{equation}
where the joint distribution of $X,A,Y$ in \eq{eq: near lossless
reconstrcution min rate} is of the form
\begin{equation}\label{eq: form of joint dist for act-dep rd func NEARLOSSLESS}
  P_{X,A,Y}(x,a,y) =  P_X (x) P_{A|X} (a|x)
  P_{Y|X,A}(y|x,a),
\end{equation}
and the minimization is over all $P_{A|X}$. Letting $R_{SW} (P_X,
P_{Y|X})$ denote the conditional entropy $H(X|Y)$ induced by the
pair $(P_X, P_{Y|X})$ (the subscript $SW$ standing for
`Slepian-Wolf' \cite{SlepianWolf73}), it is natural to wonder
whether the above minimum rate can be strictly better (smaller) than
$\min_a R_{SW} (P_X, P_{Y|X, A=a})$, which is what would be achieved
if the decoder greedily takes the one action leading to S.I.\ which
is best in the sense of inducing lowest $H(X|Y)$, irrespective of
any information from the encoder, and then proceeding as in
Slepian-Wolf coding. The following is an example showing that this
greedy strategy may be suboptimal.

\begin{figure}[h!]{
\psfrag{X}[][][1]{$X\sim \left\{ \begin{array}{cc}0 & \mbox{ w.p. }
1/2 \\ 1 & \mbox{ w.p. } 1/2 
\end{array}\right\}$}

                \psfrag{Encoder}[][][1.1]{Encoder}
\psfrag{Decoder}[][][1.1]{Decoder} \psfrag{Vender}[][][1.1]{Vender}
\psfrag{T}[][][1]{$\;\;\;\;\;\ R$} \psfrag{hat}[][][1]{$\;\hat X^n$}
\psfrag{T2}[][][1]{$A$}
 \psfrag{Y}[][][1]{$Y$}  \psfrag{XX}[][][1]{$X$}
\psfrag{T5}[][][1]{$$} \psfrag{T6}[][][1]{$$}
 \psfrag{1}[][][0.8]{$1$}
\psfrag{-1}[][][0.8]{$-1$}
 \psfrag{0}[][][0.8]{$0$}
\psfrag{a}[][][0.7]{$1-\delta$} \psfrag{b}[][][0.7]{$\delta$}
\psfrag{ZZ}[][][1]{Z channel} \psfrag{SS}[][][1]{S channel}

\centerline{\includegraphics[width=10cm]{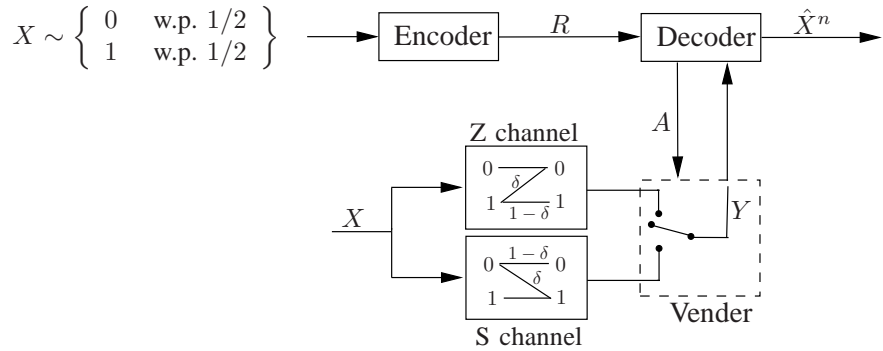}}
\caption{An example of vending side information, where the action
chooses between Z-channel and S-channel with parameter $\delta$.}
\label{f_zchannel} }\end{figure}

\begin{figure}[h!]{

\psfrag{D}[][][1]{$D$} \psfrag{z^n}[][][1]{$z^n$}
\psfrag{xhatn}[][][1]{$\hat{x}^n$}

\psfrag{0.4}[][][1]{$\begin{array}{c} \\  0.4 \\
\;\;\;\;\;\;\;\;\;\;\;\;\;\; \delta
\end{array}$}
\psfrag{0.2}[][][1]{$\begin{array}{c} \\  0.2 \\ {}
 \end{array}$}
\psfrag{0.6}[][][1]{$\begin{array}{c} \\  0.6 \\ {}
 \end{array}$}
\psfrag{0.8}[][][1]{$\begin{array}{c} \\  0.8 \\ {}
 \end{array}$}
\psfrag{1}[][][1]{$\begin{array}{c} \\  1 \\ {}
 \end{array}$}

\psfrag{0.006}[][][1]{$R_{greedy} (\delta) - R_{min}
(\delta)\;\;\;\;\;\;
0.006$\;\;\;\;\;\;\;\;\;\;\;\;\;\;\;\;\;\;\;\;\;\;\;\;\;\;\;\;\;\;\;\;\;\;\;\;\;\;}
\psfrag{0.008}[][][1]{$\;0.008$} \psfrag{0.01}[][][1]{$\;0.01$}
\psfrag{0.004}[][][1]{$\;0.004$} \psfrag{0.002}[][][1]{$\;0.002$}

\psfrag{Encoder}[][][1]{Encoder} \psfrag{Decoder}[][][1]{Decoder}
\psfrag{Unknown Source}[][][1]{Unknown Source}
\psfrag{Unknown}[][][1]{Unknown} \psfrag{Source}[][][1]{Source}
\psfrag{General}[][][1]{General} \psfrag{channel}[][][1]{channel}
\psfrag{Denoiser}[][][1]{Denoiser}
\psfrag{Systematic}[][][1]{Systematic}
\psfrag{Encoder}[][][1]{Encoder} \psfrag{Known}[][][1]{Known}
\psfrag{Denoiser -}[][][1]{Denoiser -} \psfrag{DMC}[][][1]{DMC}
\psfrag{Decoder}[][][1]{Decoder}
\centerline{\includegraphics[width=8cm]{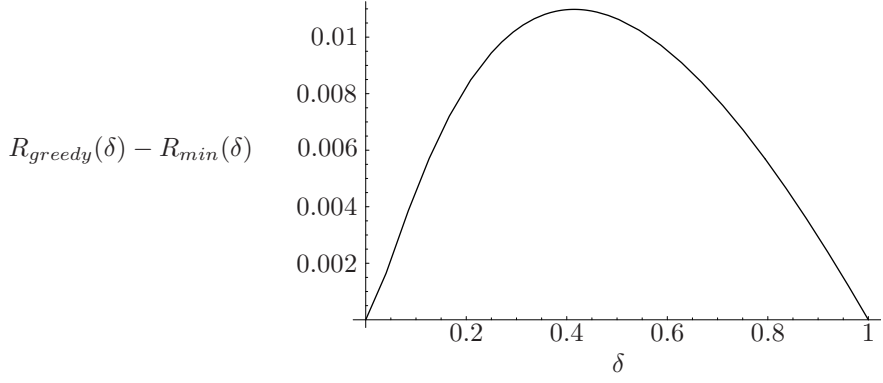}}
\caption{Plot of $R_{greedy} (\delta) - R_{min} (\delta)$}
\label{fig: rgreedy minus rmin}
}\end{figure}

Consider the case $\mathcal{X} = \mathcal{A} = \mathcal{Y} = \{
0,1\}$ where $X$ is a fair coin flip, $P_{Y|X,A=0}$ is the Z-channel
with crossover probability $\delta$ from $1$ to $0$, and
$P_{Y|X,A=1}$ is the S-channel with crossover probability $\delta$
from $0$ to $1$. The setting is depicted in Figure \ref{f_zchannel}.
Symmetry implies that the $P_{A|X}$ minimizing $I(X;A) + H(X|Y,A)$
satisfies $P_{A|X} (0|1) = P_{A|X} (1|0)$, in other words, there is
a BSC connecting $X$ to $A$ (or $A$ to $X$). Assuming this BSC has
crossover probability $\alpha$, an elementary calculation yields
\begin{equation}\label{eq: elementary calc for binary}
    I(X;A) + H(X|Y,A) = 1 - h(\alpha) + h \left( \frac{\alpha \delta}{1 - \alpha + \alpha \delta}
    \right) (1 - \alpha + \alpha \delta).
\end{equation}
Thus, letting $R_{min} (\delta)$ denote the minimum in \eq{eq: near
lossless reconstrcution min rate} for this scenario,
\begin{equation}\label{eq: rmin for the zchannel}
R_{min} (\delta) = \min_{\alpha \in [0,1]} \left[ 1 - h(\alpha) + h
\left( \frac{\alpha \delta}{1 - \alpha + \alpha \delta}
    \right) (1 - \alpha + \alpha \delta) \right].
\end{equation}
In contrast, the minimum rate achieved by a `greedy' strategy which
chooses actions without regard to the information from the encoder
is given by the conditional entropy of the input given the output of
the Z-channel($\delta$) whose input is a fair coin flip, namely
\begin{equation}\label{eq: rgreedy for z  channel}
R_{greedy} (\delta) = h \left( \frac{\delta}{1 + \delta} \right)
\frac{1 + \delta}{2}.
\end{equation}
For example, elementary calculus shows that $R_{min} (1/2)$ is
achieved by $\alpha^* = 2/5$, assuming the value $\approx 0.678072$,
which is about a $1.5\%$ improvement over $R_{greedy} (1/2) \approx
0.688722$. Figure  \ref{fig: rgreedy minus rmin} plots the
difference between $R_{greedy} (\delta)$ and  $R_{min} (\delta)$.

\begin{figure}[h!]{

\psfrag{0.2}[][][1]{$\begin{array}{c} \\  0.2 \\
\;\;\;\;\;\;\;\;\;\;\;\;\;\; C
\end{array}$}
\psfrag{0.4}[][][1]{$\begin{array}{c} \\  0.4 \\ {}
 \end{array}$}
\psfrag{0.1}[][][1]{$\begin{array}{c} \\  0.1 \\ {}
 \end{array}$}
\psfrag{0.3}[][][1]{$\begin{array}{c} \\  0.3 \\ {}
 \end{array}$}
\psfrag{0.5}[][][1]{$\begin{array}{c} \\  0.5 \\ {}
 \end{array}$}

\psfrag{0.684}[][][1]{$R_{min} (1/2, C)\;\;\;\;\;\;
0.684$\;\;\;\;\;\;\;\;\;\;\;\;\;\;\;\;\;\;\;\;\;\;\;\;\;}
\psfrag{0.686}[][][1]{$\;\;0.686$} \psfrag{0.01}[][][1]{$\;0.01$}
\psfrag{0.004}[][][1]{$\;0.004$} \psfrag{0.002}[][][1]{$\;0.002$}

 \psfrag{0.688}[][][1]{$\begin{array}{c} R_{greedy}(1/2)\to\;\;\;\;\;\;\;\;\;\;\;\;\;\\{}\\ \end{array}$} \psfrag{0.678}[][][1]{
$R_{min}(1/2)\;\;\;\;\;\;\;\;\;$} \psfrag{z^n}[][][1]{$z^n$}
\psfrag{xhatn}[][][1]{$\hat{x}^n$}
 \psfrag{DMC}[][][1]{DMC}
\psfrag{Encoder}[][][1]{Encoder} \psfrag{Decoder}[][][1]{Decoder}
\psfrag{Unknown Source}[][][1]{Unknown Source}
\psfrag{Unknown}[][][1]{Unknown} \psfrag{Source}[][][1]{Source}
\psfrag{General}[][][1]{General} \psfrag{channel}[][][1]{channel}
\psfrag{Denoiser}[][][1]{Denoiser}
\psfrag{Systematic}[][][1]{Systematic}
\psfrag{Encoder}[][][1]{Encoder} \psfrag{Known}[][][1]{Known}
\psfrag{Denoiser -}[][][1]{Denoiser -} \psfrag{DMC}[][][1]{DMC}
\psfrag{Decoder}[][][1]{Decoder}
\centerline{\includegraphics[width=8cm]{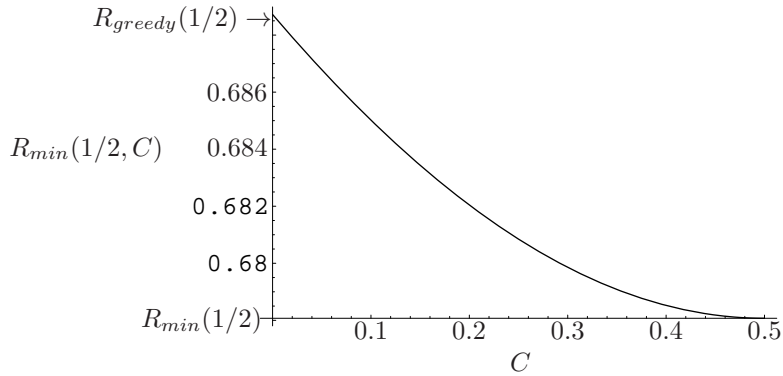}}
\caption{Plot of $R_{min} (1/2, C)$ as a function of the cost $C$.}
\label{fig: from rgreedy to rmin of half as func of cost}
}\end{figure}

In the presence of a cost constraint, Theorem   \ref{th: main
theorem for si vending at decoder} implies that the minimum rate
needed for a near lossless reconstruction is given by the minimum in
\eq{eq: near lossless reconstrcution min rate}, with the additional
constraint $E \Lambda (A) \leq C$. Let $R_{min} (\delta, C)$ denote
this minimum for our present example, assuming cost $0$ for using
say the first Z-channel and $1$ for using the second channel.
Clearly $R_{min} (\delta, 0) = R_{greedy} (\delta)$,  $R_{min}
(\delta, 1/2) = R_{min} (\delta)$ and consequently, by a
time-sharing argument,
\begin{equation}\label{eq: time sharing for lossless example with cost}
    R_{min} (\delta, C) \leq 2 C R_{min}
(\delta) + (1-2 C) R_{greedy} (\delta) \ \ \ \ 0 \leq C \leq 1/2.
\end{equation}
As it turns out, the inequality in \eq{eq: time sharing for lossless
example with cost} is strict, i.e., in our example one can do better
than time-sharing between the respective optimum schemes for the
different costs (to the level allowed by the cost constraint).
Figure  \ref{fig: from rgreedy to rmin of half as func of cost}
contains a plot of $R_{min} ( 1/2, C)$, which is seen to be better
(lower) than the straight line represented by the right side of
\eq{eq: time sharing for lossless example with cost}.

\subsubsection{The Lossy Case}

\emph{Ternary Source and Binary Side Information of Unit Cost:}
Consider a ternary $X$ taking values in  $\{-1, 0, 1\}$, distributed
according to
\begin{equation}\label{eq: distribution of ternary X}
    X = \left\{ \begin{array}{cc}
                  1 & \mbox{ w.p. } 1/4 \\
                  0 & \mbox{ w.p. } 1/2 \\
                  -1 & \mbox{ w.p. } 1/4.
                \end{array}
     \right.
\end{equation}

\begin{figure}[h!]{
\psfrag{X}[][][1]{$X\sim \left\{ \begin{array}{cc}1 & \mbox{ w.p. }
1/4 \\ 0 & \mbox{ w.p. } 1/2 \\-1 & \mbox{ w.p. } 1/4
\end{array}\right\}$}

                \psfrag{Encoder}[][][1.1]{Encoder}
\psfrag{Decoder}[][][1.1]{Decoder} \psfrag{Vender}[][][1.1]{Vender}
\psfrag{T}[][][1]{$\;\;\;\;\;\ R$} \psfrag{hat}[][][1]{$\;\hat X^n$}
\psfrag{T2}[][][1]{$A$}
 \psfrag{Y}[][][1]{$Y$}  \psfrag{XX}[][][1]{$X$}
\psfrag{T5}[][][1]{$$} \psfrag{T6}[][][1]{$$}
 \psfrag{1}[][][0.8]{$1$}
\psfrag{-1}[][][0.8]{$-1$}
 \psfrag{0}[][][0.8]{$0$}
\psfrag{a}[][][0.6][-45]{$0.5$} \psfrag{b}[][][0.6][45]{$0.5$}

\centerline{\includegraphics[width=10cm]{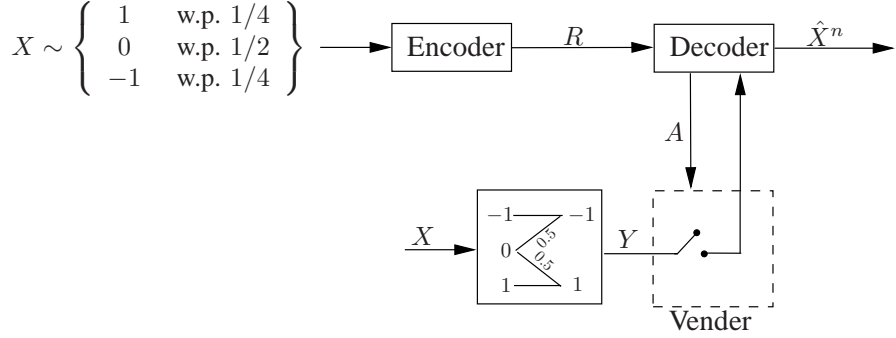}}
\caption{Ternary example.}  \label{f_binary} }\end{figure}

The actions are binary, taking values in $\{0,1\}$, where action $0$
corresponds to no S.I.\ while action $1$ corresponds to obtaining a
binary noisy measurement of $X$, taking values in $\{-1, 1\}$, which
is the output of the following channel: $P_{Y|X} (1|1) = P_{Y|X}
(-1|-1) = 1$ and $P_{Y|X} (1|0) = P_{Y|X} (-1|0) = 1/2$. Suppose
that there is a unit cost for obtaining such a noisy measurement of
the source, i.e.: $\Lambda (a) = a$, $a \in \{0,1\}$. 

 The conditional entropy of $X$ given $Y$  is $1$ bit. Thus,
lossless compression of $X$ is achievable at a rate of $1$ bit per
source symbol at a cost of $1$ per source symbol with a greedy
decoder who chooses to observe the noisy source measurement of all
symbols. Can one do better than this greedy policy? This rate is
achievable at half the cost via the following scheme: the encoder
uses one bit per source symbol to describe whether or not the symbol
is $0$. The decoder then needs to use the noisy measurement of the
source only for those symbols that are not $0$ (in which case the
measurement will completely determine the source symbol). This
corresponds to rate $I(X ; A) + H(X  | Y, A)$ under $P_{A|X}(1|1) =
P_{A|X}(1|-1) = P_{A|X}(0|0)=1$, which is readily verified to be the
minimum of achievable rates under a cost constraint of $1/2$.

In the lossy case, under Hamming distortion, we note that:
\begin{itemize}
  \item When the S.I.\ is available to both encoder and decoder (at no
  cost) the problem is reduced to one of lossy compression for the
  binary symmetric source, thus $R_{X|Y} (D) = 1 - h(D)$.

  \item This rate is achievable even when the S.I.\ is absent at the
  encoder, as can be seen by letting $W$ be the output of a BSC($D$)
  whose input  $Q(X)$ is the quantized version of $X$, defined by  $Q(0)=0$ and
  $Q(1) = Q(-1) = 1$, where $W-X-Y$. It is readily seen that the
  optimal estimate of $X$ based on  $(W,Y)$ satisfies $P(X \neq \hat{X}(W,Y)) =
  D$ and that $I(X;W|Y) = 1 - h(D)$. Thus $R_{X|Y}^{WZ} (D) = R_{X|Y} (D) = 1 -
  h(D)$.

  \item $R_{X|Y}^{WZ} (D)$ in the above item corresponds to a
  decoder that observes all of the S.I. symbols. Can the same performance be achieved with fewer observations? In other words,
  assuming unit cost per observation, can the same performance be
  achieved at a cost less than $1$? We now argue that the same
  performance can be achieved at half the cost: letting, as before,
  $A=0$ correspond to no observation and $A=1$ correspond to an
  observation, consider a conditional distribution $P_{A|X}$ given
  by $P_{A|X}( 1|1 )= P_{A|X}( 1|-1 ) = 1- P_{A|X}( 1|0) = D$ and
  where $U=A$. Then $I(X ; A) + I(X ; U | Y, A) = I(X ; A) = H(A) - H(A|X) = 1 -
  h(D)$ and the optimal estimate of $X$ based on $(U,Y)$ has $P(X \neq \hat{X} (U,Y)) =
  D$. The cost here is $P(A=1) = 1/2$. Evidently, the
  rate-distortion-cost  function $R(D, C)$ in  \eq{eq: rd function for action dependent si}
   satisfies $R(D, 1/2) \leq 1 - h(D)$ and in fact $R(D, 1/2) = 1 -
  h(D)$ since obviously $R(D, 1/2) \geq R_{X|Y}^{WZ} (D)$. Thus the
  rate $1 - h(D)$ is achievable even if the decoder is allowed to
  access only half of the observations.

\end{itemize}

\subsubsection{Binary Action: To Observe or Not to Observe the
S.I.} Consider a given source and side information distribution
$P_{X,Y}$. The action is to either observe the side information
symbol or not, where an observation has unit cost. Thus  $0 \leq C
\leq 1$ is a constraint on the fraction of side information symbols
the decoder will be allowed to observe. Let us arbitrarily take
$\mathcal{A} = \{0,1\}$, with $A=1$ corresponding to observation of
the side-information symbol and $A=0$ to lack of it. Noting that the
second mutual information term in \eq{eq: rd function for action
dependent si} corresponds to Wyner-Ziv coding conditional on $A$,
the specialization of Theorem \ref{th: main theorem for si vending
at decoder} for this case gives
\begin{eqnarray}
   & & R(D,C) \nonumber \\ & = & \min_{P_{A|X}: A-X-Y, P(A=1) = C, (1-C) D_0 + C D_1 =D } I(X;A) + R(P_{X|A=0}, D_0) \cdot P(A=0)
    + R_{WZ}(P_{X,Y|A=1}, D_1) \cdot P(A=1) \nonumber \\ & = & \min_{P_{A|X}: A-X-Y, P(A=1) = C, (1-C) D_0 + C D_1 =D } I(X;A) +
    R(P_{X|A=0}, D_0) \cdot (1- C)
    + R_{WZ}(P_{X,Y|A=1}, D_1) \cdot C \label{eq: rdc when to observe or not},
\end{eqnarray}
where $R(P_X, D)$ denotes the rate distortion function of the source
$P_X$ and $R_{WZ}(P_{X,Y}, D)$ denotes the Wyner-Ziv rate distortion
function when source and side information are distributed according
to $P_{X,Y}$.

A very special case is when $Y=X$. Thus the action is either to
observe the source symbol or not. Assuming a non-negative distortion
measure satisfying $\min_{\hat{x}} \rho (x , \hat{x}) = 0$ for all
$x$, \eq{eq: rdc when to observe or not} becomes
\begin{equation}\label{eq: rdc when si is x}
R(D,C) = \min_{P_{A|X}:  P(A=1) = C} I(X;A) +
    R \left( P_{X|A=0}, \frac{D}{1-C} \right) \cdot (1- C).
\end{equation}
When $X$ is a fair coin flip and distortion is Hamming, \eq{eq: rdc
when si is x} becomes (for $D,C$ in the non-trivial region)
\begin{eqnarray}
   & & R(D,C) \nonumber \\ & = & \min_{P_{A|X}:  P(A=1) = C} I(X;A) +
    R_b \left( P_{X=1|A=0}, \frac{D}{1-C} \right) \cdot (1- C) \nonumber \\ & = & \min_{P_{A|X}:  P(A=1) = C} 1 - [ h_b (P_{X=1|A=1}) C +
    h_b (P_{X=1|A=0}) (1-C)] +  h_b \left( P_{X=1|A=0} \right)- h_b \left( \frac{D}{1-C} \right)
 \cdot (1- C), \nonumber \\  & = & \min_{P_{A|X}:  P(A=1) = C} 1 -  h_b (P_{X=1|A=1}) C  - h_b \left( \frac{D}{1-C} \right)
 \cdot (1- C), \nonumber \\ & \stackrel{(a)}{=} &  1  -C  - h_b \left( \frac{D}{1-C} \right)
 \cdot (1- C), \nonumber \\ & = & R_b \left( \frac{1}{2} ,  \frac{D}{1-C}  \right) \cdot (1-
 C) \nonumber \\ & = & R_{A \perp  X} (D, C)
\end{eqnarray}
where $R_b (p, D) = [ h_b(p)- h_b(D)] ^+$ is the rate distortion
function of the Bernoulli($p$) source and step (a) is due to the fact that $-  h_b
(P_{X=1|A=1})$ is minimized (at the value $-1$) by taking $A$
independent of $X$.

To see that $R(D,C)$ can be strictly smaller than $R_{A \perp X} (D,
C)$ in the observe/not-observe binary action scenario, consider the
case where $X$ is a fair coin flip and $Y$ is the output of an
erasure channel with erasure probability $\sf e$ (whose input is
$X$). Recalling that $R_{WZ}(P_{X,Y}, D) = {\sf e} R(P_X, D/{\sf
e})$ when $Y$ is the erased version of
 $X$ (cf.\ \cite{{verduweissmanerasure08}, {PDT}}), we specialize the right hand side of \eq{eq: rdc when to observe or
not} for this case to obtain
\begin{eqnarray} \label{eq: rdc when to observe or not binary and erasure}
   & &  R(D,C) \nonumber \\ & = & \min_{P_{A|X}: A-X-Y, P(A=1) = C, (1-C) D_0 + C D_1 =D } 1-H(X|A) + R_b (P_{X|A=0}(1), D_0) \cdot
    (1-C)
    + {\sf e} R_{b}(P_{X|A=1}(1), D_1/{\sf e}) \cdot C \nonumber \\
    & = & \min 1 - \left[ h_b \left( \frac{\beta}{2 (1 - C )}
    \right) (1-C) + h_b \left( \frac{1-\beta}{2 C} \right) C \right]
    + R_b \left( \frac{\beta}{2 (1 - C )}, \frac{D-C D_1}{1-C} \right)
    (1-C)
    + {\sf e} R_{b}\left( \frac{1-\beta}{2 C} , D_1/{\sf e} \right)
    C  , \nonumber \\ \label{eq: the rdc for binary with erasure and
    cost}
\end{eqnarray}
where the last minimum is over $\max \{ 0 , 1 - 2C \} \leq \beta
\leq \min \{1, 2 - 2 C \}$ and $0 \leq D_1 \leq \min \{ D/C, \sf e
\}$. For the extreme points we get, as expected: $R(D,0) = R_b
\left( \frac{1}{2 }, D \right)$ and $R(D,1) = {\sf e} R_{b}\left(
\frac{1}{2} , D/{\sf e} \right)$. Figure \ref{fig:
binarysource_with_erasure_toseeornot} plots the curve in \eq{eq: the
rdc for binary with erasure and cost} for $D=1/4$, $\sf e = 1/2$ and
$0 \leq C \leq 1$.
\begin{figure}[h!]{

\psfrag{0.4}[][][1]{$\begin{array}{c} \\  0.4 \\
\;\;\;\;\;\;\;\;\;\;\;\;\;\;\;\;\;\; C
\end{array}$}
\psfrag{0.2}[][][1]{$\begin{array}{c} \\  0.2 \\ {}
 \end{array}$}
\psfrag{0.6}[][][1]{$\begin{array}{c} \\  0.6 \\ {}
 \end{array}$}
\psfrag{0.8}[][][1]{$\begin{array}{c} \\  0.8 \\ {}
 \end{array}$}
\psfrag{1}[][][1]{$\begin{array}{c} \\  1 \\ {}
 \end{array}$}

\psfrag{0.1}[][][1]{$R \left( D=\frac{1}{4},C
\right)\;\;\;\;\;\;\;\;\;\;\;\;
0.1$\;\;\;\;\;\;\;\;\;\;\;\;\;\;\;\;\;\;\;\;\;\;\;\;\;\;\;\;\;\;}
\psfrag{0.125}[][][1]{$\;\;\;\;\;0.125$}
\psfrag{0.075}[][][1]{$\;\;\;\;\;0.075$}
\psfrag{0.025}[][][1]{$\;\;\;\;\;0.025$}
\psfrag{0.05}[][][1]{$\;\;\;\;0.05$}
\psfrag{0.15}[][][1]{$\;\;\;\;0.15$}
\psfrag{0.175}[][][1]{$\;\;\;\;\;0.175$}

\psfrag{Encoder}[][][1.1]{Encoder}
\psfrag{Decoder}[][][1.1]{Decoder} \psfrag{Vender}[][][1.1]{Vender}
\psfrag{T}[][][1]{$\;\;\;\;\;\;\;\;\;T(X^n)\in 2^{nR}$}
\psfrag{hat}[][][1]{$\;\hat X_i(Y^i,T)$}
\psfrag{T2}[][][1]{$A_i(T)\;\;\;\;$}
 \psfrag{Y}[][][1]{$\;\;\;\;\; \;\;\; \;\;\;\;\; \;\;\; Y_i\sim P_{Y_i|A_i,X_i}$}
\psfrag{T5}[][][1]{$$} \psfrag{T6}[][][1]{$$}

\centerline{\includegraphics[width=10cm]{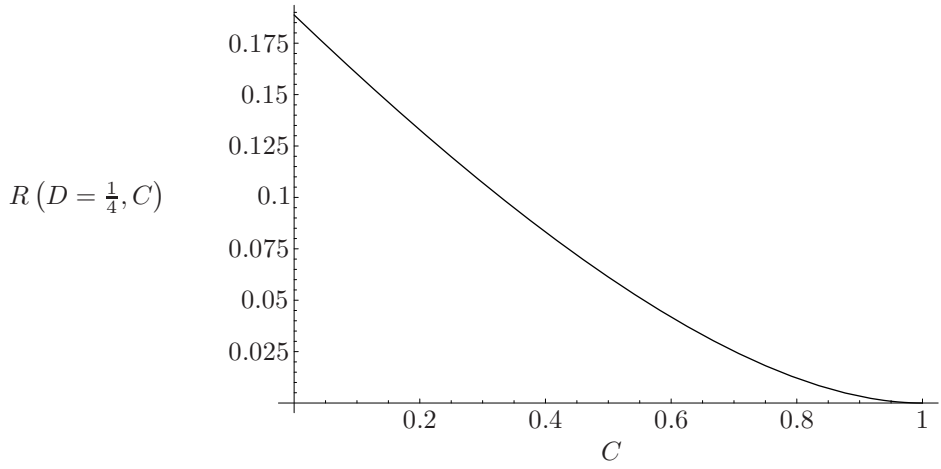}}

\caption{Rate distortion cost function $R(D,C)$, $0 \leq C \leq 1$,
for the case where $X$ is a fair coin flip, $Y$ its erased version
where $\sf e = 1/2$, $C$ is fraction of places where decoder is
allowed to observe S.I., and $D=1/4$. In this case $R(D,0) = R_b
\left( \frac{1}{2 }, \frac{1}{4} \right) \approx 0.188722$, $R(D,1)
= {\sf e} R_{b}\left( \frac{1}{2} , D/{\sf e} \right) = \frac{1}{2}
R_{b}\left( \frac{1}{2} , \frac{1}{2} \right) = 0$. The strict
concavity implies sub-optimality of time-sharing optimal schemes
according to the available observation budget.} \label{fig:
binarysource_with_erasure_toseeornot} }\end{figure}

\subsection{Causal Decoder Side Information} 
\begin{figure}[h!]{
\psfrag{X}[][][1]{$X^n$} \psfrag{Encoder}[][][1.1]{Encoder}
\psfrag{Decoder}[][][1.1]{Decoder} \psfrag{Vender}[][][1.1]{Vender}
\psfrag{T}[][][1]{$\;\;\;\;\;\;\;\;\;T(X^n)\in 2^{nR}$}
\psfrag{hat}[][][1]{$\;\hat X_i(Y^i,T)$}
\psfrag{T2}[][][1]{$A_i(T)\;\;\;\;$}
 \psfrag{Y}[][][1]{$\;\;\;\;\; \;\;\; \;\;\;\;\; \;\;\; Y_i\sim P_{Y_i|A_i,X_i}$}
\psfrag{T5}[][][1]{$$} \psfrag{T6}[][][1]{$$}

\centerline{\includegraphics[width=10cm]{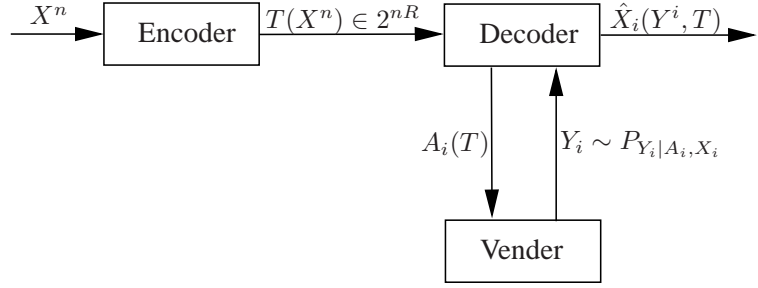}}

\caption{Rate distortion with {\it causal} side information vender
at the decoder.} \label{f_decoder_vender_causal} }\end{figure}

Consider the setting presented in
Figure \ref{f_decoder_vender_causal}, which is similar to that
described in Section \ref{sec: the setup}, the only difference being
that the reconstruction is allowed  causal dependence on  the side
information, i.e., to be of the form $\hat{X}_i = \hat{X}_i (T,
Y^i)$ (motivation for why this might be interesting can be found in
\cite{withAbbas06}).

Define
\begin{equation}\label{eq: rd function causal}
    R^{(I)}_{causal}(D, C) = \min I(X ; U,A) ,
\end{equation}
where the joint distributions of $X,A,Y,U$ is of the form
\begin{equation}\label{eq: form of joint dist for act-dep rd func again}
  P_{X,A,U,Y}(x,a,u,y) =  P_X (x) P_{A,U|X} (a,u|x)
  P_{Y|X,A}(y|x,a),
\end{equation}
and the minimization is over all $P_{A,U|X}$ under which
\begin{equation}\label{eq: the expected distortion constraint}
    E \left[ \rho \left( X, \hat{X}^{opt} (U,Y) \right) \right] \leq
    D,  \ \ \ \ E \left[  \Lambda (A) \right]
\leq C,
\end{equation}
where $\hat{X}^{opt} (U,Y)$ denotes the best estimate of $X$ based
on $U,Y$, where $U$ is an auxiliary random variable. The cardinality
of $U$ may be restricted to $|\mathcal{U}| \leq |\mathcal
X||\mathcal A|+2$ as shown in item \ref{it: cardinality} Lemma
\ref{lemma:properties_R}. One can also denote $U,A$ as $\tilde U$,
and an equivalent representation would be
\begin{equation}\label{eq: rd function causal}
    R^{(I)}_{causal}(D, C) = \min I(X ; \tilde U) ,
\end{equation}
where $P_{X,A,\tilde U,Y}(x,a,\tilde u,y) =  P_X (x) P_{\tilde U|X}
(\tilde u|x)1_{\{ a = f(\tilde u) \}}
  P_{Y|X,A}(y|x,a)$.

\begin{theorem}
The rate distortion cost function for the setting where actions
taken by the decoder before the index is seen, is given by
$R^{(I)}_{causal}(D, C)$.
\end{theorem}
\begin{proof}
{\bf Achievability:} The achievability proof is based on the fact
that the encoder and decoder generate a joint type $P_{X,A}$ using a
rate that is $I(X;A)+\epsilon$, and since both the encoder and
decoder know the sequence of actions $a^n$, they can time-share
between $|\mathcal A|$ causal schemes such that if the action is $a$
a rate $I(X;U|a)+\epsilon$ would achieve the distortion constraint
\cite{withAbbas06}. Hence, the total rate is $I(X;A)+\epsilon+\sum
P_A(a)(I(X;U|a)+\epsilon)=I(X;A,U)+2\epsilon$.

{\bf Converse:} for the converse part, fix a scheme of rate $R$ for
a block of length $n$ and consider:
\begin{eqnarray}\label{eq: conv1}
nR&\stackrel {\geq}{}& H(T)\nonumber \\
&\stackrel {\geq}{}& I(X^n;T)\nonumber \\
&\stackrel {=}{}& \sum_{i=1}^n H(X_i) -H(X_i|X^{i-1},T)\nonumber \\
&\stackrel {(a)}{=}& \sum_{i=1}^n H(X_i) -H(X_i|X^{i-1},T,Y^{i-1})\nonumber \\
&\stackrel {}{\geq}& \sum_{i=1}^n H(X_i) -H(X_i|T,Y^{i-1})
\end{eqnarray}
where step (a)  is due to the Markov chain
$X_i-(X^{i-1},T)-Y^{i-1}.$ Now let us denote $\tilde
U_i:=(T,Y^{i-1})$, and we obtain that
\begin{equation}
R\geq \frac{1}{n}I(X_i;\tilde U_i).
\end{equation}

The proof is now completed in the standard way upon considering the
joint distribution of $(X',A',\tilde U', Y', \hat{X}') \mug (X_J,
A_J, (\tilde U_J,J) , Y_J, \hat{X}_J)$, where $J$ is randomly
generated uniformly at random from the set $\{1, \ldots, n\}$,
independent of $(X^n, A^n, U^n, Y^n, \hat{X}^n)$, and noting that:

  \begin{equation}\label{eq: distribution properties again}
   P_{X'} = P_X,\ \tilde U' - (A',X')-Y',\ P_{Y'|X',A'} =
  P_{Y|X,A},
  \end{equation}
  \begin{equation}
   \hat{X}' = \hat{X}' (\tilde U', Y'), \ A'=f(\tilde U'),
  \end{equation}
\begin{equation}\label{eq: distortion and cost constraint single let}
    E \left[ \sum_{i=1}^n \rho (X_i, \hat{X}_i) \right] = n E \rho
    (X', \hat{X}')
, \ \ \ \
    E \left[ \sum_{i=1}^n \Lambda (A_i) \right] = n E \Lambda (A')
\end{equation}
   and
\begin{equation}\label{eq: convexity of mut inf for sum vs generic}
     \frac{1}{n} \sum_{i=1}^n  I (X_i ; \tilde U_i) = I(X'; \tilde
     U').
\end{equation}

 \end{proof}

\subsection{Indirect Rate Distortion with Action-Dependent Side
Information}

\begin{figure}[h!]{
\psfrag{X}[][][1]{$X^n$} \psfrag{Z}[][][1]{$Z^n$}
\psfrag{Encoder}[][][1.1]{Encoder}
\psfrag{Decoder}[][][1.1]{Decoder} \psfrag{Vender}[][][1.1]{Vender}
\psfrag{T}[][][1]{$\;\;\;\;\;\;\;\;\;T(Z^n)\in 2^{nR}$}
\psfrag{hat}[][][1]{$\;\;\hat X^n(Y^n,T)$}
\psfrag{T2}[][][1]{$A^n(T)\;\;\;\;$}
 \psfrag{Y}[][][1]{$Y^n$}
  \psfrag{Pzx}[][][1]{$P_{Z|X}$}
\psfrag{T5}[][][1]{$$} \psfrag{T6}[][][1]{$$}


\centerline{\includegraphics[width=14cm]{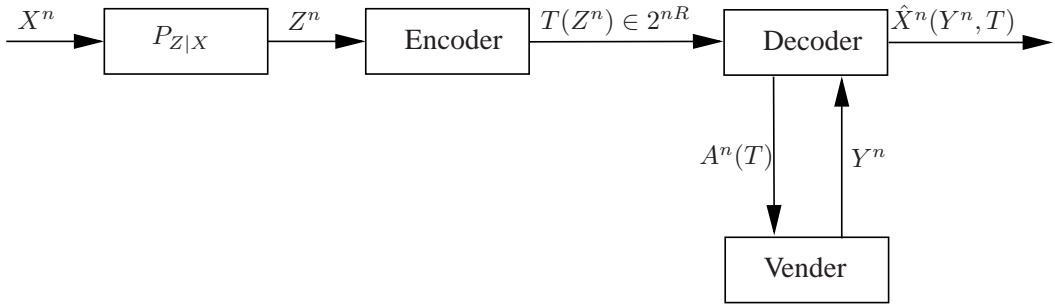}}

\caption{Indirect rate distortion with a side information vender at
the decoder. The source $X^n$ is i.i.d. $P_X$ and the encoder
observes a noisy version of the source, $Z^n$, where the pairs
$(X_i, Z_i)$ are i.i.d.$\sim P_{X,Z}$.  Side information is
generated as the output of the channel $P_{Y|X,Z,A}$ in response to
the noise-free,  noisy, and action sequences $(X^n, Z^n, A^n)$,
where the action sequence $A^n$ is generated on the basis of the
index from the encoder.} \label{f_decoder_vender_inderect}
}\end{figure}

Consider the case shown in Figure \ref{f_decoder_vender_inderect}
where, rather than the source $X$, the encoder observes a noisy
version of it, $Z$. The decoder, based on the index conveyed to it
from the encoder,  will then select an action sequence that will
result in the side information $Y$, as output from the channel
$P_{Y|X,Z,A}$. The reconstruction, as before, will be a function of
the index and the side information. Specifically, a scheme in this
setting for blocklength $n$ and rate $R$ is characterized by an
encoding function $T : \mathcal{Z}^n \rightarrow \{1,2, \ldots,
2^{nR} \}$, an action strategy $f: \{1,2, \ldots, 2^{nR} \}
\rightarrow \mathcal{A}^n$, and a decoding function $g : \{1,2,
\ldots, 2^{nR} \} \times \mathcal{Y}^n \rightarrow
\hat{\mathcal{X}}^n$ that operate as follows:
\begin{itemize}
  \item The source $n$-tuple $X^n$ is i.i.d.$\sim P_X$ goes through a
  DMC   $P_{Z|X}$ to yield its noisy observation sequence $Z^n$.
  Thus, overall the clean and noisy source are characterized by a
  given joint distribution $P_{X,Z}$

  \item Encoding: based on $Z^n$ give index $T = T(Z^n)$ to the decoder

  \item Decoding:
  \begin{itemize}
    \item given the index, choose an action sequence $A^n = f(T)$

    \item   the side information
    $Y^n$ will be the output of the memoryless channel $P_{Y|X,Z,A}$
    whose input is $(X^n,  Z^n, A^n)$

    \item let $\hat{X}^n = g (T, Y^n)$
  \end{itemize}
\end{itemize}
The rate-distortion-cost for this case is now defined similarly as
in subsection \ref{sec: the setup}. Let us denote it by
$R_{ID}(D,C)$, the subscript standing for `indirect'. Theorem
\ref{th: main theorem for si vending at decoder} is generalized to
this case as follows:
\begin{theorem} \label{th: main theorem for si vending at decoder indirect}
$R_{ID}(D,C)$ is given by
\begin{equation}\label{eq: rdc func equals operational indirect}
R_{ID}(D,C) = \min \left[ I(Z ; A) + I(Z ; U | Y, A) \right] ,
\end{equation}
where the joint distribution of $X, Z, A,Y,U$  is of the form
\begin{equation}\label{eq: form of joint dist for act-dep rd func indir}
  P_{X, Z, A,U,Y}(x, z, a,u,y) =  P_{X,Z} (x,z) P_{A,U|Z} (a,u|z)
  P_{Y|X,Z,A}(y|x, z, a),
\end{equation}
and the minimization is over all $P_{A,U|Z}$ under which
\begin{equation}\label{eq: the expected distortion constraint}
    E \left[ \rho \left( X, \hat{X}^{opt} (U,Y) \right) \right] \leq
    D,  \ \ \ \ E \left[  \Lambda (A) \right]
\leq C,
\end{equation}
where $\hat{X}^{opt} (U,Y)$ denotes the best estimate of $X$ based
on $U,Y$, and $U$ is an auxiliary random variable whose cardinality
is bounded as $|\mathcal{U}| \leq |\mathcal Z||\mathcal A|+2$.
\end{theorem}
\emph{Proof outline:} The achievability part is very similar to the
original. The random generation of the scheme is performed in the
same way, with the noisy source replacing the original noise-free
source. This guarantees that $(Z^n, A^n, U^n, Y^n)$ are, with high
probability, jointly typical. The joint typicality also with $X^n$,
namely the joint typicality of $(X^n, Z^n, A^n, U^n, Y^n)$, then
follows from an application of the Markov lemma. The converse part
also follows similarly to the one from the noise-free case: that
\begin{equation}\label{eq: same as noisefree part i}
     n R  \geq  I(Z^n ; A^n) + H(Z^n|A^n,
    Y^n) - H(Z^n|A^n,Y^n,T)
\end{equation}
follows identically as in \eq{eq: converse chain of ineq} by
replacing $X^n$ by $Z^n$. That
\begin{equation}\label{eq: same as noisefree part ii}
    I(Z^n ; A^n) + H(Z^n|A^n, Y^n) \geq \sum_{i=1}^n I(Z_i; A_i) + H (Z_i| Y_i, A_i)
\end{equation}
follows similarly as \eq{eq: converse chain of ineq part2} by
replacing $X^n$ with $Z^n$, upon noting that $H(Y^n | A^n, Z^n) =
\sum_{i=1}^n H(Y_i | A_i, Z_i)$, which follows from the Markov
relation $(X_i, Y_i) - (A_i, Z_i) - (A^{n \setminus i}, Z^{n
\setminus i}, Y^{i-1} )$ (which a fortiori implies $Y_i - (A_i, Z_i)
- (A^{n \setminus i}, Z^{n \setminus i}, Y^{i-1} )$). Combining
\eq{eq: same as noisefree part i} and \eq{eq: same as noisefree part
ii} now yields
\begin{equation}\label{eq: now with the aux for indirect}
    n R  \geq  \sum_{i=1}^n I(Z_i; A_i) + I (X_i ; U_i | Y_i, A_i)
\end{equation}
similarly as in Step (a) in \eq{eq: defining aux U} upon defining  $U_i=(A^{n
\setminus i}, Y^{n \setminus i}, Z^{i-1},T)$. The proof of the
converse is concluded by verifying that:
\begin{itemize}
  \item $\hat{X}_i
= \hat{X}_i (T, Y^n)$ is a function of the pair $(U_i,Y_i)$
  \item the
Markov relation $X_i - Z_i -(A_i, U_i)$ holds (which follows from
$X_i - Z_i - (Z^n , Y^{n \setminus i})$)
  \item the
Markov relation $U_i - (X_i, Z_i, A_i) - Y_i$ holds (which follows
from $(Z^n , Y^{n \setminus i}) - (X_i, Z_i, A_i)- Y_i$)
\end{itemize}
and invoking the convexity of the informational rate distortion
function defined on the right hand side of \eq{eq: rdc func equals
operational indirect}, which is established similarly as in Lemma
\ref{lemma:properties_R}. \hfill \QED

\section{Side Information Vending Machine at the Encoder}
\label{sec: Side Information Vending Machine at the Encoder} In this
section we consider the setting where the action sequence $A^n$ is
chosen at the encoder and the side information is available at the
decoder and possibly at the encoder too. The setting is depicted in
Figure \ref{f_action_encoder_switch}. Specifically, a communication
scheme in this setting for blocklength $n$ and rate $R$ is
characterized by an action strategy
\begin{equation}f: \mathcal{X}^n \rightarrow
\mathcal{A}^n,\end{equation} an encoding function
\begin{eqnarray} T &:&
\mathcal{X}^n\times \mathcal{Y}^n \rightarrow \{1,2, \ldots, 2^{nR}
\}\; \text{(when side information is available at the encoder),}
\nonumber \\
 T &:&\mathcal{X}^n \rightarrow \{1,2, \ldots, 2^{nR} \} \;\;\;\;\;\;\;\;\;\; \text{(when
side information is {\bf not} available at the encoder),} \nonumber
\end{eqnarray}
 and a decoding function
\begin{equation}g : \{1,2, \ldots, 2^{nR} \} \times \mathcal{Y}^n
\rightarrow \hat{\mathcal{X}}^n.\end{equation}

As in the case where the actions were chosen by the decoder, the
side information $Y^n$ will be the output of the memoryless channel
$P_{Y|X,A}$ whose input is $(X^n, A^n)$. Furthermore, a triple
$(R,D,C)$ is said to be \emph{achievable} if for all $\eps
>0$ and sufficiently large $n$ there exists a scheme as above for blocklength $n$ and rate
$R + \eps$ satisfying both
\begin{equation}\label{eq: distortion constraint}
    E \left[ \sum_{i=1}^n \rho (X_i, \hat{X}_i) \right] \leq n (D+\eps)
\end{equation}
and
\begin{equation}\label{eq: cost constraint}
    E \left[ \sum_{i=1}^n \Lambda (A_i) \right] \leq n (C+\eps) .
\end{equation}
The rate distortion (and cost) function $R_e(D,C)$ (The letter $e$
stands for encoder) is defined as
\begin{equation}\label{eq: rate distortion function defined operationally atencoder}
R_e(D,C) = \inf \{ R' : \mbox{ the triple } (R',D,C) \mbox{ is
achievable} \}.
\end{equation}


The general case remains open, however we present here a
characterization of three important cases: lossless case (where
$\Pr(X^n=\hat X^n)\to 1$), Gaussian case (where $Y=A+X+N$ and $X$
and $N$ are independent Gaussian random variables), and a case where
the Markov form $Y-A-X$ holds. In all three cases, $R_e(D,C)$ is
independent of whether or not the S.I.\ is available at the encoder.

\subsection{Lossless case} \label{subsec: Lossless case}
Here we consider the  lossless case, namely, for any $\epsilon>0$
there exists an $n$ such that $\Pr(X^n=\hat X^n)>1-\epsilon.$ Define
\begin{equation}\label{eq:R_lossles_encoder}
    R_e^{(I)}(C) = \min \left[ H(X| A,Y) + I(X ; A)-I(Y;A) \right],
\end{equation}
 where $P_X$ and
$P_{Y|A,X}$ are determined by the problem setting and the
minimization is over $P_{A|X}$ such that $E \left[ \Lambda (A)
\right] \leq  C$. The term $\left[ H(X| A,Y) + I(X ; A)-I(Y;A)
\right]$ is convex in $P_{A|X}$ since the term $-I(Y;A,X)$ is convex
in $P_{A|X}$  and the following identity holds
\begin{eqnarray}
H(X|A,Y)+I(X;A)-I(Y;A)&=& H(X|A,Y)+H(X)-H(X|A)-H(Y)+H(Y|A)\nonumber
\\
&=& H(X) -I(X;Y|A)-H(Y)+H(Y|A)\nonumber\\
&=& H(X) -H(Y|A)+H(Y|A,X)-H(Y)+H(Y|A)\nonumber\\
&=& H(X) -I(Y;A,X)
\end{eqnarray}

Let us denote the minimum (operational) rate that is needed to
reconstruct the source at the encoder losslessy where with a cost of
the action less than $C$ as $R_e(C)$.

\begin{theorem}
For the setting in Figure \ref{f_action_encoder_switch} where the
actions are chosen by the encoder and the side information $Y$ is
known to the decoder and may or may not be known to the encoder the
minimum rate that is needed to reconstruct the source under a cost
constraint $C$ is given by
\begin{equation}\label{eq:R_lossles_encoder}
    R_e^{}(C) = R_e^{(I)}(C).
\end{equation}
\begin{proof}

{\bf Achievability :} The achievability proof is divided into two
cases according to the sign of the term  $I(X;A)-I(Y;A)$. In the
first case we assume $I(X;A)-I(Y;A)> 0$ and we use a coding scheme
that is based on Wyner-Ziv coding \cite{WynerZiv1976} for rate
distortion theory where side information known at the decoder. In
the second case, we assume $I(Y;A)-I(X;A)> 0$ and we use a coding
scheme that is based on Gel'fand-Pinsker coding \cite{GePi80} for
channel with states where the state is known to the encoder.

{\it First case $I(X;A)-I(Y;A)> 0$ :} We first generate a codebook
of sequences of actions $A^n$ that covers $X^n$; hence,  the size of
the codebook needs to be  $2^{n(I(A;X)+\epsilon)}$, where
$\epsilon>0$. Then, similarly to Wyner-Ziv coding scheme
\cite{WynerZiv1976}, we bin the codebook into
$2^{n(I(A;X)-I(Y;A)+2\epsilon)}$ bins such that into each bin we
have $2^{n(I(Y;A)-\epsilon)}$ codebooks. Similarly to Wyner-Ziv
scheme, we look in the codebook for a sequence $A^n$ that is jointly
typical with $X^n$ and transmit the number of the bin that contains
the sequence.  The decoder receives the bin number and looks which
of the sequences of $A^n$ in the bin that its number is received are
jointly typical with the side information $Y^n$. Similar to the
analysis in Wyner-ziv setting, with high probability there will be
only one codeword that is jointly typical with $Y^n$ (The Markov
form that is needed in the analysis of Wyner-ziv setting is not
needed here, since the side information $Y^n$ is generated according
to $P_{Y|A,X}$ and therefore if $(A^n,X^n)$ are jointly typical then
with high probability the triple $(A^n,X^n,Y^n)$ would also be
jointly typical). In the final step the encoder uses a Slepian-Wolf
scheme for transmitting $X^n$ losslessy to the decoder that has side
information $(Y^n,A^n)$; hence additional rate of $H(X|Y,A)$ is
needed.

{\it Second case $I(Y;A)-I(X;A)> 0$ :} First we notice that the
expression in (\ref{eq:R_lossles_encoder}) can be written as  $H(X|
A,Y) -(I(Y;A)- I(X ; A))$. The actions can be considered as input to
a channel with states where the output of the channels is $Y$ and
the state is $X$ and the conditional probability of the channel is
$P_{Y|X,A}$. The capacity of this channel is achieved by
Gel'fand-Pinsker coding scheme \cite{GePi80} and is given as
$I(Y;A)- I(X ; A)$. In addition the Gel'fand-Pinsker coding scheme
induces a triple $(X^n,Y^n,A^n)$ that is jointly typical. Hence, we
can use the message in order to reduce the needed rate $H(X|Y,A)$ as
in Slepian-Wolf scheme to $H(X| A,Y) -(I(Y;A)- I(X ; A))$.

{\bf Converse:} for the converse part, fix a scheme of rate $R$ for
a block of length $n$ with a probability of error $\Pr(X^n\neq
X^n)=P_e^{(n)}$ and consider:
\begin{eqnarray}\label{eq: conv1}
nR&\stackrel {\geq}{}& H(T)\nonumber \\
&\stackrel {\geq}{}& H(T|Y^n) \nonumber \\
&\stackrel {=}{}& H(X^n,T|Y^n)-H(X^n|T,Y^n)\nonumber \\
&\stackrel {(a)}{\geq}& H(X^n,T|Y^n)- n\epsilon_n\nonumber \\
&\stackrel {(b)}{=}& H(X^n,A^n|Y^n)- n\epsilon_n\nonumber \\
&\stackrel {}{=}& H(X^n,A^n,Y^n)-H(Y^n)- n\epsilon_n\nonumber \\
&\stackrel {}{=}& H(X^n)+H(A^n|X^n)+H(Y^n|A^n,X^n)-H(Y^n)-
n\epsilon_n \nonumber \\
&\stackrel {(c)}{\geq}& \sum_{i=1}^n H(X_i)+H(Y_i|A_i,X_i)-H(Y_i)-
n\epsilon_n\nonumber \\
 &\stackrel {}{\geq}& \min\left [n
\left(H(X)+H(Y|A,X)-H(Y)\right) \right]- n\epsilon_n
\end{eqnarray}
where $\epsilon_n=P_e^{(n)}\log|\mathcal X|+\frac{1}{n}$ and step
(a) follows Fano's inequality. Step (b) follows the fact that $A^n$
and $T$ are deterministic functions of $X^n$. Step (c) follows the
following four relations: $P(x^n)=\prod_{i=1}^nP(x_i)$,
$P(y^n|a^n,x^n)=\prod_{i=1}^nP(y_i|a_i,x_i)$   $H(A^n|X^n)=0$, and
$H(Y^n)\leq \sum_{i=1}^n H(Y_i)$. The minimization in the last step
is over all conditional distribution ${P_{A|X}}$ that satisfy the
cost constrain, namely  $E \left[\Lambda (A) \right] \leq C$,  and
the inequality follows from the fact that the expression
$H(X)+H(Y|A,X)-H(Y)$ is convex in $P_{A|X}$ for fixed $P_X$ and
$P_{Y|A,X}$. The converse proof is completed by invoking the fact
that since $R$ is an achievable rate there exists a sequence of
codes at rate $R$ such that $\epsilon_n\to 0$.
\end{proof}

\end{theorem}

 We have seen that, in the
absence of a cost constraint on the actions, the minimum rate needed
for a near lossless reconstruction at  the decoder is  given by
\begin{equation}\label{eq: near lossless reconstrcution min rate actions@encoder}
    \min I(X ; A) + H(X  | Y, A) - I(Y;A)
\end{equation}
(regardless of whether or not the side information is present at the
encoder). Thus, $I(Y;A)$ represents the saving in rate relative to
the case where the actions are taken by the  decoder (recall \eq{eq:
near lossless reconstrcution min rate} for the minimum rate at that
case). To see that this can be significant, recall the example
$\mathcal{X} = \mathcal{A} = \mathcal{Y} = \{ 0,1\}$, where $X$ is a
fair coin flip, $P_{Y|X,A=0}$ is the Z-channel with crossover
probability $\delta$ from $1$ to $0$, and $P_{Y|X,A=1}$ is the
S-channel with crossover probability $\delta$ from $0$ to $1$. It is
easily seen that in this case $I(X ; A) + H(X  | Y, A) - I(Y;A) = 0$
and so, a fortiori, the minimum in \eq{eq: near lossless
reconstrcution min rate actions@encoder} is zero. That the source
can be reconstructed losslessly with zero rate in this  case is
equally easy to see from an operational standpoint, since taking
actions $A_i = X_i$ ensures that $Y_i = X_i$ with probability one.

\subsection{Gaussian Case} \label{subsec: Gaussian Case}
Here we consider the case where
\begin{itemize}
\item the source has a Gaussian distribution with zero mean and
variance $\sigma^2_X$, i.e.,
\begin{equation}
X\sim \text{N}(0,\sigma^2_X),
\end{equation}
\item the relation between $Y,X,A$ is given by
\begin{equation}
Y=X+A+N,
\end{equation}
where $N$ is a random variable independent of $(A,X)$ and has a
Gaussian distribution with zero mean and variance $\sigma^2_N$,
i.e.,
\begin{equation}
W\sim \text{N}(0,\sigma^2_N),
\end{equation}
\item the distortion is a mean square error distortion, i,e,
$E\left[\sum_{i=1}^n(X_i-\hat X_i)^2\right]$ and it has to be less
than $D$
\item  the  cost of the actions is $E\left[
\sum_{i=1}^nA_i^2\right]$ and  has to be less than $C$. Without loss
of generality, we assume that $C=\alpha^2  \sigma_X^2$ where
$\alpha>0$.
\end{itemize}
\begin{theorem}\label{t_gaussian_action_encoder}
For the Gaussian setting of Figure \ref{f_action_encoder_switch}, as
described above,
\begin{equation}\label{e_gaussian_encoder_action}
R_e(D,C)= \left\{ \begin{array}{cc}
                    \frac{1}{2}\log \left[ \frac{\sigma^2_N
}{(1+ \sqrt{C}/\sigma_X )^2\sigma^2_X+\sigma^2_N} \cdot \frac{\sigma^2_X}{D} \right] & \mbox{ if }
[(1+ \sqrt{C}/\sigma_X )^2\sigma^2_X+\sigma^2_N] \cdot D \leq \sigma^2_X \sigma^2_N  \\
                    0  & \mbox{ otherwise.}
                  \end{array}
  \right.
\end{equation}
\end{theorem}
%
%

\begin{figure}[h!]{
\psfrag{R}[][][0.9]{{ $ R(D,C)$}} \psfrag{D}[][][0.8]{ $D$}
\psfrag{C1}[][][0.7]{\color{blue} $R(D,C=0) $}
\psfrag{C2}[][][0.7]{\color{green} $R(D,C=0.3) $}
\psfrag{C3}[][][0.7]{\color{black} $R(D,C=0.6) $}
\psfrag{C4}[][][0.7]{\color{red} $R(D,C=1) $}

\centerline{\includegraphics[width=7cm]{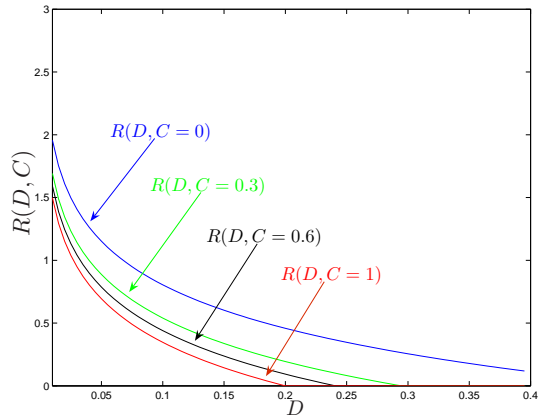}}
\caption{$R_e(D,C)$ in the Gaussian case, for $\sigma_X^2 =
\sigma_N^2=1$. The boundary of the region where $R_e(D,C) = 0$ is
the curve $D = \frac{1}{(1 + \sqrt{C})^2+1}$. Indeed, this
distortion level can be achieved with zero rate by estimating $X$ on
the basis of $Y=X+N/(1+\sqrt{C})$.} \label{fig: gaussian example}
}\end{figure}

Before proving Theorem \ref{t_gaussian_action_encoder} we would like
to point out that the state amplification problem
\cite{KimSutivongCover08StateAmplification,
SutivongCoverChiang08TradeoffStateMessage} is tangent to the vending
side information problem described here. In the state amplification
problem, the goal is to design a communication scheme for a channel
with i.i.d. states sequence, $S^n$, which is known to the encoder.
The purpose of the scheme is  to send a message through the channel,
and at the same time to describe to the decoder the state sequence
$S^n$. The case where there is no message to send, namely, the input
to the channel is used only to describe the state sequence, is
equivalent to the problem presented here when $R_e(D,C)=0$, namely,
we when are using only the actions to describe the source and no
additional message is sent. If $R_e(D,C)=0$, we obtain from
(\ref{e_gaussian_encoder_action}) that for the Gaussian source
coding problem,  the minimum mean square error satisfies
\begin{equation}
D\geq \frac{\sigma_X^2\sigma_N^2}{(\sigma_X+\sqrt{C})^2+\sigma_N^2},
\end{equation}
a result that was also obtained in \cite[Theorem
2]{SutivongCoverChiang08TradeoffStateMessage}, where the channel is
the Gaussian channel and the goal is to describe the state sequence
with minimum mean square error distortion.

{\it Proof of Theorem \ref{t_gaussian_action_encoder}: }

{\bf Achievability:} The encoder chooses the actions to be $A=\alpha
X$ and then it uses a coding for the Gaussian Wyner-Ziv
 with side information at the decoder
\cite{Wyner78_gaussian_WZ}. The side information satisfies
$Y=X+A+N=(1+\alpha)X+N$, which is equivalent to having a side
information $Y=X+\frac{N}{(1+\alpha)}$. Denote by
$N'=\frac{N}{1+\alpha}$. Using the Gaussian Wyner-Ziv result, a rate
\begin{equation}
R=\frac{1}{2}\log\frac{\sigma^2_{N'}}{\sigma^2_X+\sigma^2_{N'}}\frac{\sigma^2_X}{D}=\frac{1}{2}\log
\frac{\sigma^2_N }{(1+\alpha)^2\sigma^2_X+\sigma^2_N}
\frac{\sigma^2_X}{D},
\end{equation}
 is achievable.

{\bf Converse:} We prove the converse in two steps. First we derive
the lower bound
\begin{equation}
R_e(D,C)\geq \min_{P_{A|X},P_{\hat X|X,A,Y}} I(X;\hat X)-I(Y;X,A),
\end{equation}
which holds for any $P_X$ and $P_{Y|A,X}$ (not necessarily
Gaussian), and then we evaluate it for the Gaussian case.

Fix a scheme at rate $R$ for a block of length $n$ and consider
\begin{eqnarray}
nR&\geq& H(T) \nonumber \\
&\geq& H(T|Y^n) \nonumber \\
&\geq& I(X^n;T|Y^n) \nonumber \\
&\stackrel{(a)}{=}& I(X^n;Y^n,T)-I(X^n;Y^n) \nonumber \\
&\stackrel{(b)}{\geq}& I(X^n;\hat X^n)-I(X^n,A^n;Y^n) \nonumber \\
&\stackrel{(c)}{\geq}& \sum_{i=1}^n I(X_i;\hat
X_i)-I(X_i,A_i;Y_i),\label{e_con_gen}
\end{eqnarray}
where (a) follows from
\cite[Lemma3.2]{Wyner78_conditional_mutual_info}, which asserts that
for arbitrary random variables $I(X,Z;Y)=I(Z;Y)+I(X;Y|Z)$. Step (b)
follows from the facts that $\hat X^n$ is a determinstic function of
the pair $(T,Y^n)$, and $A^n$ is a deterministic function of $X^n$.
Step (c) follows from the facts that $H(X^n)=\sum_{i=1}^nH(X_i)$,
$H(Y^n|A^n,X^n)=\sum_{i=1}^nH(Y_i|A_i,X_i)$ and conditioning reduces
entropy. Since the expression in (\ref{e_con_gen}) is convex in
$P_{A|X},P_{\hat X|X}$ for fixed $P_X$ and $P_{Y|A,X}$, we obtain
the lower bound
\begin{equation}\label{e_lower_general_bound_action_enc}
R_e(D,C)\geq \min I(X;\hat X)-I(Y;X,A),
\end{equation}
and the minimization is over conditional distributions
$P_{A|X},P_{\hat X|X}$ that satisfy the distortion and cost
constraints.

Now we evaluate the lower bound for the Gaussian case.
\begin{eqnarray}
I(X;\hat X)-I(Y;X,A)&=&H(X)-H(X|\hat X)-H(Y)+H(Y|A,X)\nonumber \\
&=&H(X)-H(X|\hat X)-H(Y)+H(N)\nonumber \\
&\stackrel{(a)}{\geq}& \frac{1}{2}\log 2\pi e \sigma^2_X-\frac{1}{2}\log 2\pi e D- \frac{1}{2}\log 2\pi e ((1+\alpha)^2\sigma^2_X+\sigma^2_N)) +\frac{1}{2}\log 2\pi e \sigma^2_N \nonumber \\
&\stackrel{}{=}& \frac{1}{2}\log
\frac{\sigma^2_N}{(1+\alpha)^2\sigma^2_X+\sigma^2_N}\frac{
\sigma^2_X}{ D},
\end{eqnarray}
where inequality (a) follows from the fact that $H(X|\hat X)\leq
H(X-\hat X)\leq\frac{1}{2}\log 2\pi e D$ (because of the constraint
that $E(X-\hat X)^2\leq D$) and $H(Y)\leq \frac{1}{2}\log 2\pi e
\sigma_Y^2$, where
\begin{eqnarray}
\sigma_Y^2&=&E[X^2]+2E[AX]+E[A^2]+E[N^2]\nonumber \\
&\leq& E[X^2]+2\sqrt{E[X^2]E[A^2]}+E[A^2]+E[N^2] \nonumber \\
&\leq&
\sigma_X^2+2\alpha\sigma_X^2+\alpha^2\sigma_X^2+\sigma_N^2\nonumber
\\
&=&\sigma_X^2(1+\alpha)^2+\sigma_N^2.
\end{eqnarray}
\hfill \QED
\subsection{Markov Form Y-A-X} \label{subsec: Markov Form Y-A-X}
Here we consider the case where the Markov form $X-A-Y$ holds.
\begin{theorem} \label{th: actions  at encoder  markov form}
The rate distortion (and cost) function $R_e(D,C)$  for the setting
in Figure \ref{f_action_encoder_switch} when
$P_{Y|A,X}(y|a,x)=P_{Y|A}(y|a)$ satisfies
\begin{equation}\label{eq:R_lossles_encoder again}
    R_e^{}(D,C) = \left[\min(I(X;\hat X)-I(A;Y))\right]^+,
\end{equation}
where $[\cdot]^+$ denotes $min\{\cdot,0\}$ and the minimization is
over joint distributions of the form $P_{A,X,Y,\hat X}(a,x,y,\hat
x)=P_X(x)P_{A|X}(a|x)P_{Y|A}(y|a)P_{\hat X|X}(\hat x|x)$ satisfying
$E \rho (X, \hat{X} ) \leq D$ and $E \Lambda (A) \leq C$.
\end{theorem}

It is interesting to note that the solution is the difference
between a rate-distortion expression $\min_{P_{\hat X|X}}I(X;\hat X)
$ and channel capacity expression $\max_{P_A}I(A;Y)$ . I.e.,
\begin{equation}\label{eq: alternative expression of rdc}
R_e^{}(D,C) = [ R(P_X , D) - Cap (P_{Y|A}, C) ]^+,
\end{equation}
where $Cap (P_{Y|A}, C)$ denotes the capacity of the channel
$P_{Y|A}$ under  a cost-constraint $C$.

{\it proof of Theorem \ref{th: actions  at encoder  markov form}:}

{\bf Achievability:} Design a regular rate distortion code, which
needs a rate larger than $I(X;\hat X)$, and then transmits part of
the rate through the channel which has an input $A$ and output $Y$.
Therefore the total rate that is needed to be transmitted through
the index $T(X^n)$ is the difference $I(X;\hat X)-I(A;Y)$.

{\bf Converse:} We invoke the lower bound given in
(\ref{e_lower_general_bound_action_enc}) and obtain
\begin{eqnarray}
R_e(D,C)&\geq&  I(X;\hat X)-I(Y;X,A)\nonumber \\
&=&  I(X;\hat X)-I(Y;A),
\end{eqnarray}
where the last equality is due to the Markov form $X-A-Y$.\hfill
\QED

\subsection{Upper and  Lower  Bounds  for the General Case} \label{subsec: Upper and  Lower  Bounds  for the General Case}
\subsubsection{Achievable Rates}
\begin{itemize}
  \item {\bf Absence of S.I. at Encoder:} For the setting of Figure
  \ref{f_action_encoder_switch} with an open switch, i.e., when the
  encoder has no access to the S.I., the following is an achievable
  rate:
\begin{equation}\label{eq: acheivable rate for open switch}
    I(U; X | A , Y ) + I(X;A) - I(Y;A)
\end{equation}
under any joint distribution of the form
\[  P_X (x) P_{A|X} (a|x)  P_{Y|X,A} (y|x,a)
P_{U|X,A} (u|x,a) \] such that $E \rho (X, \hat{X}_{opt} (A,Y,U))
\leq D$ and $E \Lambda (A) \leq C$. The argument for why this rate
is achievable is similar to that given in Subsection \ref{subsec:
Lossless case} for why the right side of \eq{eq:R_lossles_encoder}
is achievable, the difference being that the $H(X| A,Y)$ term in
\eq{eq:R_lossles_encoder}, corresponding to Slepian-Wolf coding of
$X^n$ conditioned on $A^n$, is replaced by $I(U; X | A , Y )$,
corresponding to Wyner-Ziv coding conditioned on $A^n$.

  \item {\bf S.I. Available at Encoder:} For the setting of Figure
  \ref{f_action_encoder_switch} with a closed switch, i.e., when the
  encoder has access to the S.I., the following is an achievable
  rate:
\begin{equation}\label{eq: acheivable rate for open switch}
    I(\hat{X}; X | A , Y ) + I(X;A) - I(Y;A)
\end{equation}
under any joint distribution of the form
\[  P_X (x) P_{A|X} (a|x)  P_{Y|X,A} (y|x,a) P_{\hat{X}|X,A,Y}
 (\hat{x}|x,a,y) \] such that $E \rho (X, \hat{X})
\leq D$ and $E \Lambda (A) \leq C$. The argument for why this rate
is achievable is similar to that  for why the right side of \eq{eq:
acheivable rate for open switch} is achievable, the difference being
that the  $I(U; X | A , Y )$ terms, corresponding to Wyner-Ziv
coding conditioned on $A^n$, is replaced by $I(\hat{X}; X | A , Y
)$, corresponding to standard rate distortion coding conditioned on
$A^n, Y^n$.

\end{itemize}

\subsubsection{Lower Bound on Achievable Rate}
As pointed out in Subsection \ref{subsec: Gaussian Case},  the proof
of the converse part of Theorem \ref{t_gaussian_action_encoder} is
valid for the general case (i.e., beyond the Gaussian scenario), and
shows that the rate needed to achieve distortion $D$ at cost $C$,
regardless of whether or not S.I.\ is available at the encoder, is
at least as large as
\begin{equation}\label{eq: lower bound on rate needed}
    I(X ; \hat{X}) - I(Y; X, A)
\end{equation}
for some joint distribution of the form
\[ P_X (x) P_{A|X} (a|x) P_{Y|X,A} (y|x,a) P_{\hat{X}|X,A,Y} (\hat{x}|x,a,y) \]
satisfying the distortion and cost constraints. It is worthwhile to
note that this rate was shown to be achievable for the three special
cases considered in the previous three subsections. Indeed, this
fact was shown explicitly for the cases of  Subsection \ref{subsec:
Gaussian Case} and Subsection \ref{subsec: Markov Form Y-A-X}, and
in the lossless case \eq{eq: lower bound on rate needed} becomes
$H(X) - I(Y; X, A) = H(X|A,Y) + I(A;X) - I(A;Y)$, which coincides
with the expression on the right hand side of
\eq{eq:R_lossles_encoder}.

To see that the lower bound in \eq{eq: lower bound on rate needed}
may not be tight in  general, even when the S.I. is available at the
encoder, consider the standard case of rate distortion coding with
S.I.\ available to both encoder and decoder. In this case  $A$ is
degenerate, so the right hand side of \eq{eq: lower bound on rate
needed} reduces to
\begin{equation}\label{eq: rhs of our lower bound on rate}
    I(X ; \hat{X}) - I(Y; X, A) = H(X|Y) - H(X|\hat{X})
\end{equation}
while the tight lower bound on the achievable rate for this scenario
is well-known to be given by
\begin{equation}\label{eq: achievable rate tight siboth sides a fegenrate}
I(X ; \hat{X} | Y) = H(X|Y) - H(X|\hat{X}, Y),
\end{equation}
which may be strictly larger  than the expression in \eq{eq: rhs of
our lower bound on rate}.

\section{Summary and Open Questions}
\label{sec: Summary and Open Questions} We have studied source
coding in the presence of side information, when the system can take
actions that affect the availability, quality, or nature of the side
information. We have given a full characterization of the
rate-distortion-cost tradeoff when the actions are taken by the
decoder. For the case where the actions are taken by the encoder, we
have characterized this tradeoff in a few important special cases,
while providing upper and lower bounds on the achievable rate for
the general case.

The most significant question left open by our work is a full
characterization of the rate-distortion-cost tradeoff for the
setting of actions taken  at the encoder (beyond the special cases
considered here), with S.I.\ that may or may not be available at the
encoder (Figure \ref{f_action_encoder_switch}). Another question
left open, for the setting of actions taken by the decoder, is
whether the rate distortion cost tradeoff can be improved when each
action is allowed to depend on the side information symbols
generated thus far, that is, when the $i$th action is allowed to be
of the form $A_i = A_i (T, Y^{i-1})$ (rather than $A_i(T)$).

\section*{Acknowledgement}
We are grateful to Paul Cuff for suggesting the title of this paper
and for helpful discussions.

\end{document}